\documentclass[12pt]{article}
\usepackage{geometry} 
\usepackage{epstopdf}
\usepackage{epsfig}
\usepackage{amsmath}
\usepackage{natbib,hyperref}         
\title{A Comparision of Three Network Portfolio Selection Methods -- Evidence from the Dow Jones}

\author{Hannah Cheng Juan Zhan$^{1}$,   William Rea$^{1}$,  and Alethea Rea$^{2}$, \\
1. Department of Economics and Finance, University of Canterbury, \\
New Zealand \\
2. Data Analysis Australia, Perth, Australia }

\begin{document}

\maketitle

\begin{abstract}

\end{abstract}
We compare three network portfolio selection methods; hierarchical
clustering trees, minimum spanning trees and neighbor-Nets, with
random and industry group selection methods on twelve years of data
from the 30 Dow Jones Industrial Average stocks from 2001 to 2013
for very small private investor sized portfolios. We find that the three 
network methods perform on par with randomly selected portfolios.


\begin{description}
\item[Keywords: ] Graph theory, hierarchical clustering trees, minimum
spanning trees, neighbor-Net, portfolio selection, diversification
\item[JEL Codes: ] G11
\end{description}


\section{Introduction}\label{sec:introduction}

Portfolio diversification is critical for risk management because
it aims to reduce
the variance of returns compared with a portfolio of a single stock (or similarly
undiversified portfolio). 
The academic literature on diversification is vast,
stretching back at least as far as  \cite{lowenfeld1909}. The modern
science of diversification is usually traced to \cite{markowitz1952}
which was expanded upon in great detail in \cite{markowitz1959}.

The literature
covers a wide range of approaches to portfolio diversification, such as;
the number of
stocks 
required to form a well diversified portfolio,
which has increased from eight stocks 
in the late 1960's \citep{Evans1968} to over 100 stocks in the late 2000's \citep{Domian2007},
what types of risks should be considered, \citep{Cont2001, 
Goyal2003, Bali2005}, 
risk factors intrinsic to each
stock \citep{Fama1992, Fama1993},
the age of the investor, \citep{Benzoni2007},
whether international diversification is beneficial, 
\citep{Jorion1985,Bai2010}, among other risks.

In recent years a significant number of papers have appeared which
apply graph theoretical methods to the study of a stock or
other financial market,
see, for example, \cite{Mantegna1999}, \cite{Onnela2003a},
\cite{Onnela2003}, \cite{Bonanno2004}, \cite{Micciche2006}, \cite{Naylor2007},
\cite{Kennet2010}, and
\cite{Djauhari2012} among others.

On the pragmatic side,
\cite{DeMiguel2009} lists 15 different methods for forming portfolios
and report results from their study which evaluated 13 of these. 
Absent among these 15 methods were any which utilized the above-mentioned
graph theory approaches. This leaves as an open question whether 
these graph theory approaches can usefully be applied to the
problem of portfolio selection. 

In one sense, the approach of \cite{markowitz1952} is optimal and
cannot be bettered in the case that either the correlations and
expected returns of the assets are not time-varying thus can be accurately
estimated from historical data or, alternatively, they can
be forecast accurately. Unfortunately, neither of these conditions
hold in real markets. Indeed, \cite{Michaud1989}
studied the limitations
of the mean-variance approach and claimed that the mean-variance 
optimizer was an ``estimation error
maximizer''. These implementation problems
have left the door open to other approaches and
hence the large literature addressing this issue.

It is well-understood that the expected returns and variances of 
the individual stocks available for selection into a portfolio
are insufficient
for making an informed decision because
selecting a portfolio also requires an understanding of the correlations
between each of the stocks.
However, the number
of correlations between stocks rises in proportion to the square of
the number of stocks meaning that for all but the smallest of stock
markets
the very large number of correlations are beyond
the human ability to comprehend them. It is precisely  in this area of
understanding the correlation structure of the market that the
graph theory methods have been usefully applied.

The goal of this paper is to compare three simple network methods --
hierarchical clustering trees (HCT), minimum spanning trees (MST) and
neighbor-Nets (NN) -- with two
simple portfolio selection methods
for small private-investor sized portfolios.
There are two motivations for looking at very small portfolios sizes.

The first is that,
despite the recommendation of authorities like \cite{Domian2007},
 \cite{Barber2008} reported that in a large sample of 
American private investors
the average stock portfolio size of individual investors was only 4.3. 
Thus there is a practical
need to find way of maximising the diversification benefits
for these investors, and network methods remain largely unexplored.
The second is that
testing the methods on small 
portfolios gives us a chance to evaluate the potential benefits
of the network methods because the larger the portfolio size, the
more closely the portfolio resembles the whole market and the less
likely
any potential benefit is to be discernible.

We apply these methods to the problem of stock selection 
confining our investible universe to the 30 stocks of the
Dow Jones Industrial Average Index. The choice of such a small 
group of stocks is motivated by the fact that stock networks are
often interpreted by eye, thus having a small number of stocks
means the networks are comparatively simple and easy to interpret.
A follow-on study of a much larger set of stocks is in progress and
we hope to be able to present the results of that study in the
near future.

%
Our primary motivation is to investigate five portfolio selection
strategies, three of which involve clustering or network algorithms.  The five
 strategies are forming portfolios by picking stocks;
\begin{enumerate}
\item at random;
\item from different industry groups;
\item from different correlation clusters identified by MSTs
\item from different correlation clusters identified by HCTs
\item from different correlation clusters identified by neighbor-Net
splits graphs.
\end{enumerate}
HCT and MST have a long history of application to the study of financial
markets. Recently
\cite{Rea2014} presented a method to visualise the correlation matrix using
nieghbor-Net networks \citep{Bryant2004}, yielding insights into the 
relationships between the stocks.

The outline of this paper is as follows; Section (\ref{sec:data})
discusses the data and methods
used in this paper, Section (\ref{sec:clusters}) discusses
identifying
the correlation clusters, 
Section (\ref{sec:results}) applies
the methods and results of the previous two sections to the
problem of forming a diversified portfolio of stocks,
and Section (\ref{sec:discussion}) contains
the discussion, our conclusions and some suggestions for directions
for future research.

\section{Data and Methods}\label{sec:data}

\subsection{Data}

\begin{figure}[ht]
  \centering
  \includegraphics[width=12cm]{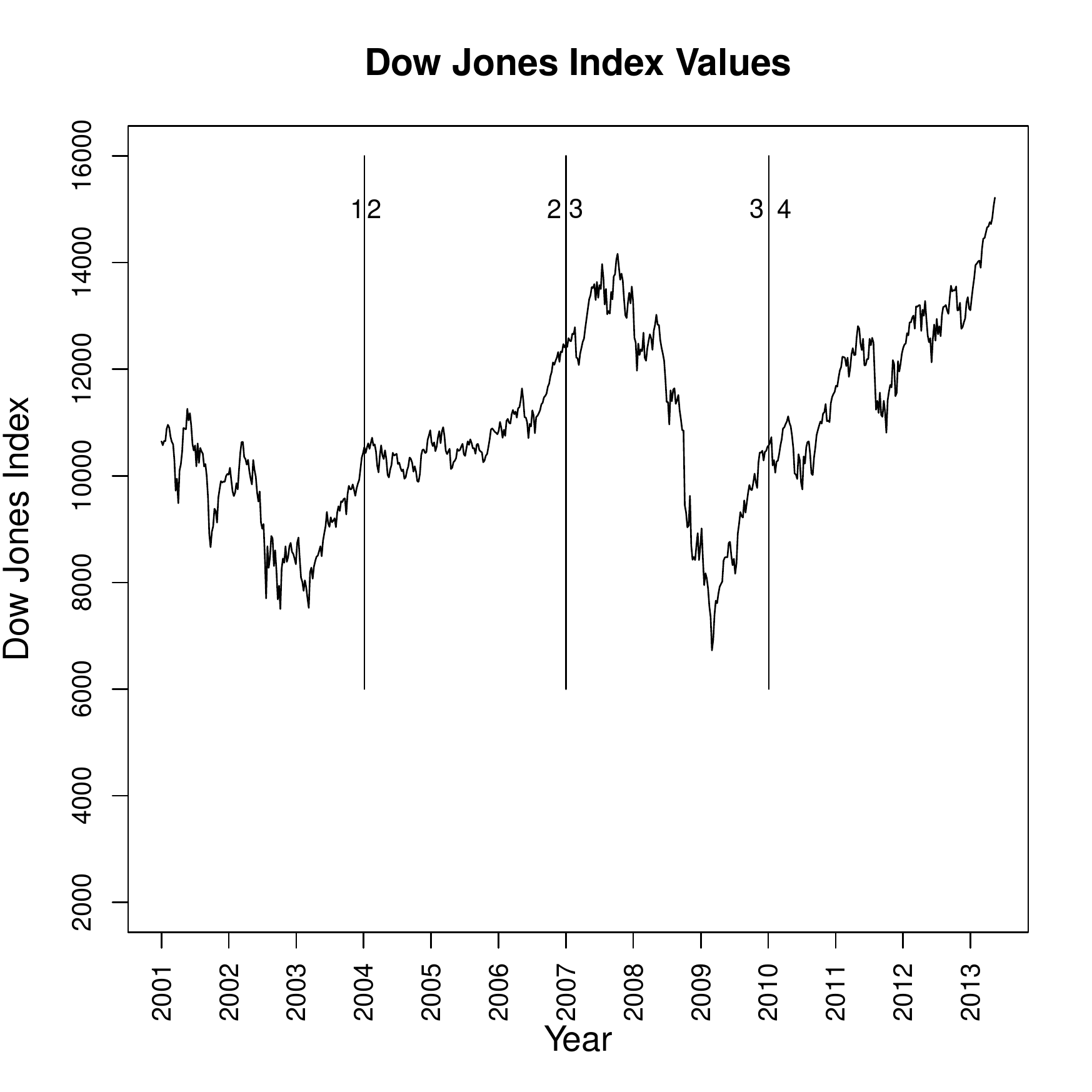}
  \caption{Dow Jones Industrial Average with the boundaries of
the four study periods marked.}
  \label{fig:DJIAIndex}
\end{figure}

We downloaded weekly closing prices along with dividend rate and dividend
payment date for each of the 30 stocks for the period
2 January 2001 to 14 May 2013 from DataStream. 
We calculated the returns by assuming that
the dividends paid were reinvested into the stock which issued them,
at the closing price on the day the dividend payment was made.

We defined four study periods, they were
\begin{enumerate}
\item 2 January 2001 to 6 January 2004.
\item 6 January 2004 to 2 January 2007.
\item 2 January 2007 to 5 January 2010.
\item 5 January 2010 to 14 May 2013.
\end{enumerate}

The first three study periods are all three years in length and are used
below for model building and in-sample testing. Periods two to four were
used for out-of-sample testing. The stocks in the sample, their ticker
symbols and industry group is given in Appendix \ref{app:stockcodes}.

Of the four study periods defined, period 3 contains the financial
crisis of 2008 and the subsequent very steep fall and rebound
in the market. Both in-sample and out-of-sample testing for period
3 represents a very severe test of a portfolio selection method.

\subsection{Clustering Diagrams}

The return series were used to generate correlation matrices using
the function \verb+cor+ in base R \citep{R}. The correlation matrices
were converted to distance matrices for use with the various clustering
algorithms using the ultra-metric
$$
d_{ij}=\sqrt{2(1-\rho_{ij})} 
$$
where $d_{ij}$ is the distance and $\rho_{ij}$ is the correlation
between stocks $i$ and $j$. For further details on the ultra-metric
see \cite{Mantegna1999}.

The hierarchical clustering trees were generated from
the distance matrix using \verb+hclust+
from the \verb+R+ \verb+stats+ package using average linkage.

The minimum spanning trees were generated using functions in the
\verb+igraph+ package \citep{igraph} within \verb+R+.

To generate the neighbor-Net splits graphs we formatted
the distance matrix and 
augmented it with the appropriate stock codes for reading into the
SplitsTree software which implements the neighbor-Net algorithm.
The neighbor-Net splits graphs were produced with \verb+SplitsTree4+ Version
4.13.1 available from \verb+http://www.splitstree.org+. 

Levene's test of equality of variances was performed with
functions implemented in the \verb+R+ package \verb+lawstat+ \citep{lawstat}.

\subsection{Simulated Portfolios}

This section describes the three methods used to simulate portfolios. 
Below results of the simulations are presented based on 
1,000 replications of each portfolio selection method.

\begin{description}
\item[Random Selection: ]
The stocks were selected at random using a uniform distribution without 
replacement. In other words each stock was given equal chance of 
being selected  but with no stock being selected twice within a 
single portfolio. 
\item[By Industry Groups: ]
In the data extracted from DataStream there were 21 different
industry groups among the 30 stocks. These were grouped into four
super-groups for the purposes of the portfolio simulations. The
original industry groups of the stocks and their assignment to
the four super-groups are listed in Appendix \ref{app:stockcodes}.

If the portfolio size was four or 
less, the industries were chosen at random using a 
uniform distribution without replacement.
From each of the selected industry groups one stock was selected. 
If the desired portfolio size was eight stocks, each group had at 
two stocks selected, again using a uniform distribution without replacement.

\item[By Correlation Clusters: ] 
There were three clustering algorithms used to assign stocks to correlation
clusters. The methods used to assign stocks to clusters is described in 
detail in Section (\ref{sec:clusters}) below and further results from 
the clustering algorithms are in Appendices 
\ref{app:Period1Graphs} and \ref{app:Period3Graphs} below.
All the portfolio simulations used the same selection method.

The correlation clusters were determined by examining the graphical
output from the clustering algorithm and stocks divided into between two and
four correlation groups based on the needs of the simulations. 
The clusters determined in periods one, two and three were used to generate 
the in-sample portfolios for the same periods and the out-of-sample
testing in periods two, three and four respectively. Because the
goal of portfolio building is to reduce risk each cluster was paired
with another cluster which was considered most distant from it where this
was feasible. With the NN splits graphs this could always be
done, with the MST groups sometimes this could be done, HCT groups 
could not be paired in this way.
\end{description}

\section{Picking Correlation Clusters}\label{sec:clusters}

\subsection{Hierarchical Clustering Trees}

\begin{figure}[ht]
  \centering
  \includegraphics[width=12cm]{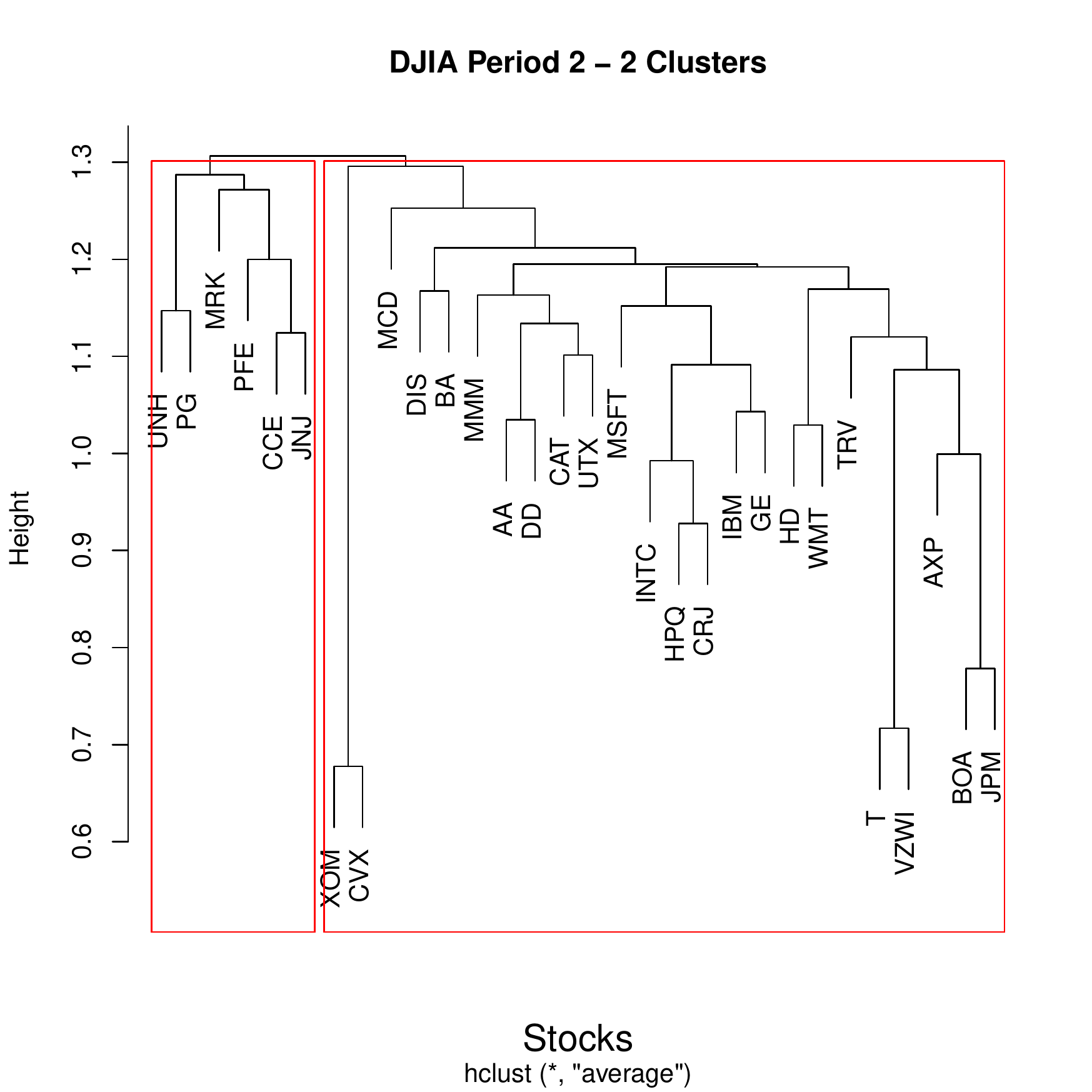}
  \caption{The HCT for Period 2 with two clusters. The unbalanced cluster
sizes are clearly seen with clusters of six and 24 stocks respectively.}
  \label{fig:DJIAHCTP42ClColoured}
\end{figure}

\begin{figure}[ht]
  \centering
  \includegraphics[width=12cm]{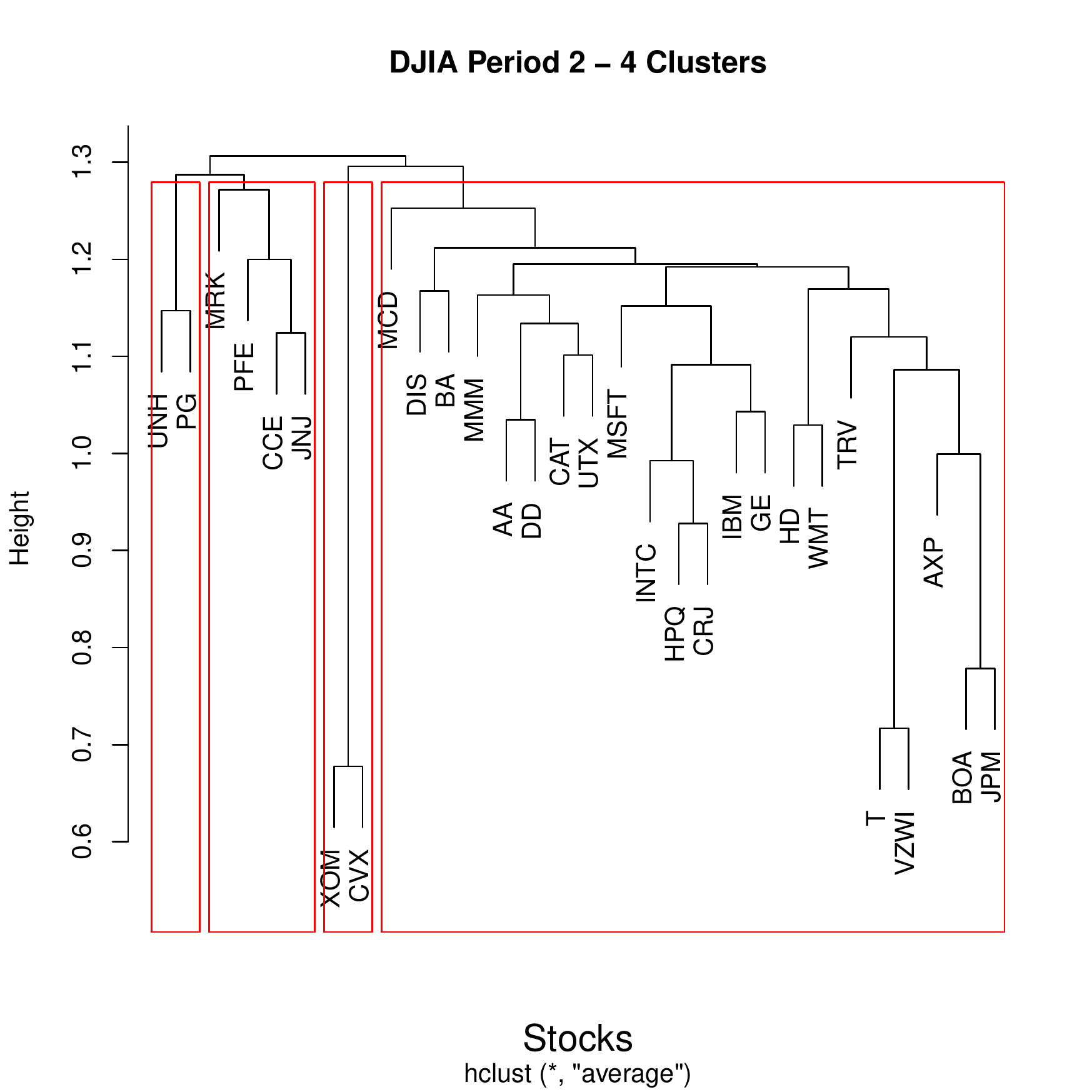}
  \caption{The HCT for Period 2 with four clusters.The unbalanced cluster
sizes are clearly seen with clusters of two, four, two and 22 stocks 
respectively.}
  \label{fig:DJIAHCTP44ClColoured}
\end{figure}

Hierarchical clustering trees are very simple to read. One starts
at the top of the tree and notes where the splits are. For example,
Figures
(\ref{fig:DJIAHCTP42ClColoured}) and (\ref{fig:DJIAHCTP44ClColoured})
show an HCT for the DJIA stocks in study period two
with two and four clusters respectively.
If one wants to divide the stocks into two clusters one simply finds
the first split and the stocks fall naturally into the two clusters desired.
If four clusters are desired, then it is simply a matter of moving
down the tree until the splits give four clusters. 

Figures
(\ref{fig:DJIAHCTP42ClColoured}) and (\ref{fig:DJIAHCTP44ClColoured})
show us a problem with HCTs which are not present to the same
extent with either MSTs or neighbor-Net splits graphs, that is,
the HCT may produce highly unbalanced clusters. In Figure
(\ref{fig:DJIAHCTP42ClColoured}) the two cluster sizes are six and
24 respectively. The problem is even worse in Figure 
 (\ref{fig:DJIAHCTP44ClColoured}) in which the cluster sizes are two,
two, four and 22. When running the portfolio simulations such
unbalanced clusters may artificially depress the standard deviations.
Given that the standard deviation is in the denominator of the
widely used Sharpe ratio \citep{Sharpe1964}, this may inflate the
apparent reward per unit risk. Thus we need to take care when interpreting
the results from the simulations when using HCTs.

A second problem with HCT clusters is that in the simulations we wished to 
pick stocks from clusters which were most distant from each other. With a
linear layout of the stocks and reading the clusters from left to right,
in a four cluster case,
pairing the left most cluster with the right most because it is most
distant, leaves the other pair adjacent to each other. Because of
this problem and the unbalanced cluster sizes, 
in the simulations we paired a large cluster with a small cluster.
Further HCTs are in Appendices \ref{app:Period1Graphs} and
\ref{app:Period3Graphs}.

\subsection{Minimum Spanning Trees}

\begin{table}
\begin{center}
\begin{tabular}{lr}
Stock & Distance \\
\hline
JPM & 0.779 \\
AXP & 0.961 \\
GE  & 1.010 \\
HD  & 1.045 \\
DD  & 1.047 \\
\hline
\end{tabular}
\end{center}
\caption{Table of Distances to decide which cluster to put BOA into for 
period 2.}\label{tab:BOADistances}
\end{table}

\begin{figure}[ht]
  \centering
  \includegraphics[width=12cm]{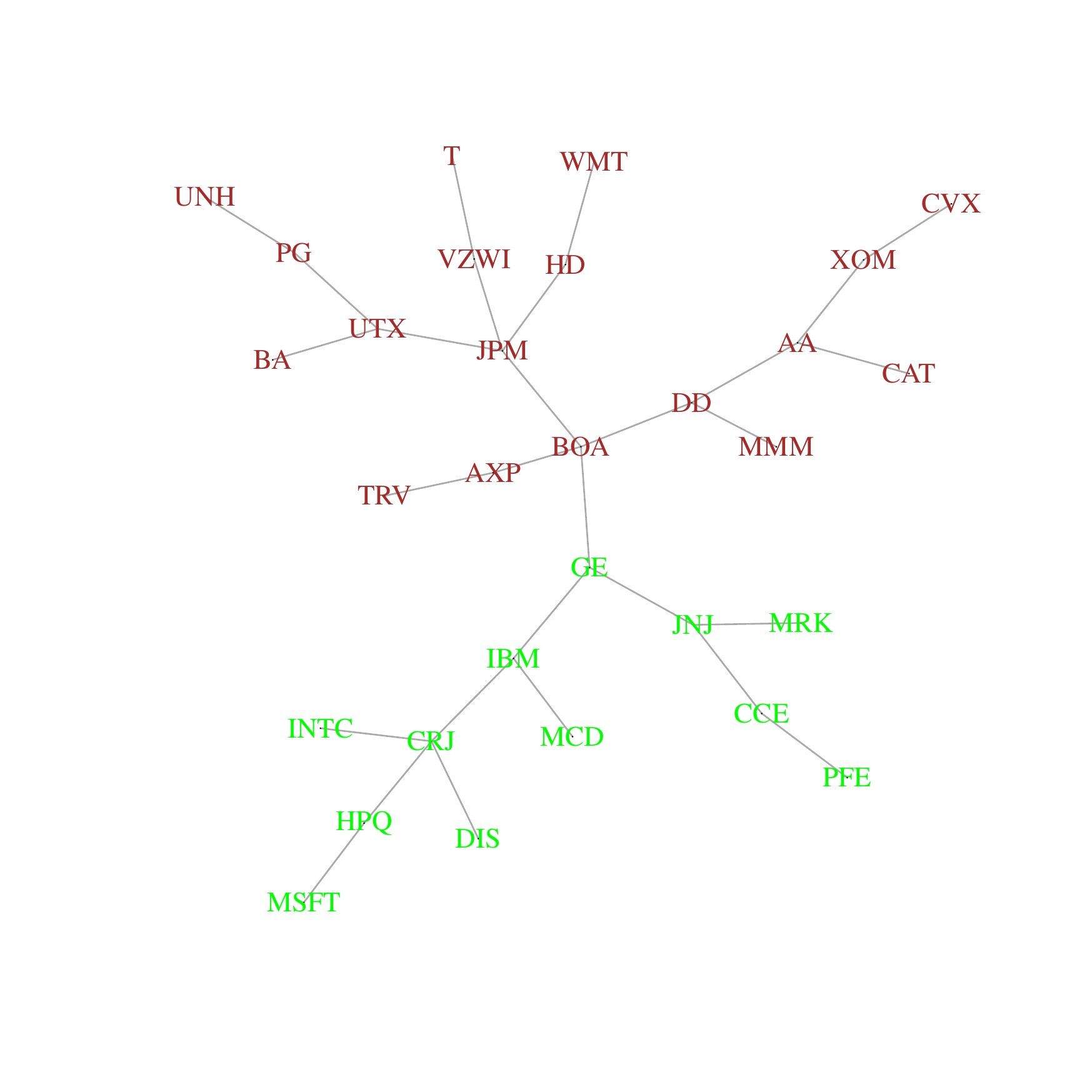}
  \caption{Minimum spanning tree for Period two DJIA with two clusters.}
  \label{fig:DJIAMSTP42ClColoured}
\end{figure}

Figure (\ref{fig:DJIAMSTP42ClColoured}) shows the MST for the DJIA for Period
2. When we divide the MST into correlation clusters, there are some
natural breaks in the tree structure which allow us to easily assign
some stocks to one particular cluster. Depending on how many clusters
we need, the tree can be divided in more than one way. If we require
two clusters it is clear that there are three distinct groupings, the
branches with roots at JPM, DD and GE respectively. The two stock branch of
TRV and AXP must go with BOA whichever cluster it is assigned to. 

In Table (\ref{fig:DJIAMSTP42ClColoured}) we have the distances
from BOA to the five nearest stocks. It is clear from this that
BOA must be assigned to the same cluster as JPM rather than the one
with GE.

\begin{table}
\begin{center}
\begin{tabular}{lr|lr}
      & Distance &      & Distance  \\
Stock & to BOA   & Stock& to AXP    \\
\hline
JPM   & 0.779    & BOA  & 0.961     \\
AXP   & 0.961    & JPM  & 1.038     \\
GE    & 1.009    & VZWI & 1.043  \\
DD    & 1.045    & GE   & 1.047 \\
\hline
\end{tabular}
\end{center}
\caption{Table of Distances to decide which cluster to put BOA and
AXP into for 
period four.}\label{tab:BOAAXPDistances}
\end{table}

\begin{figure}[ht]
  \centering
  \includegraphics[width=12cm]{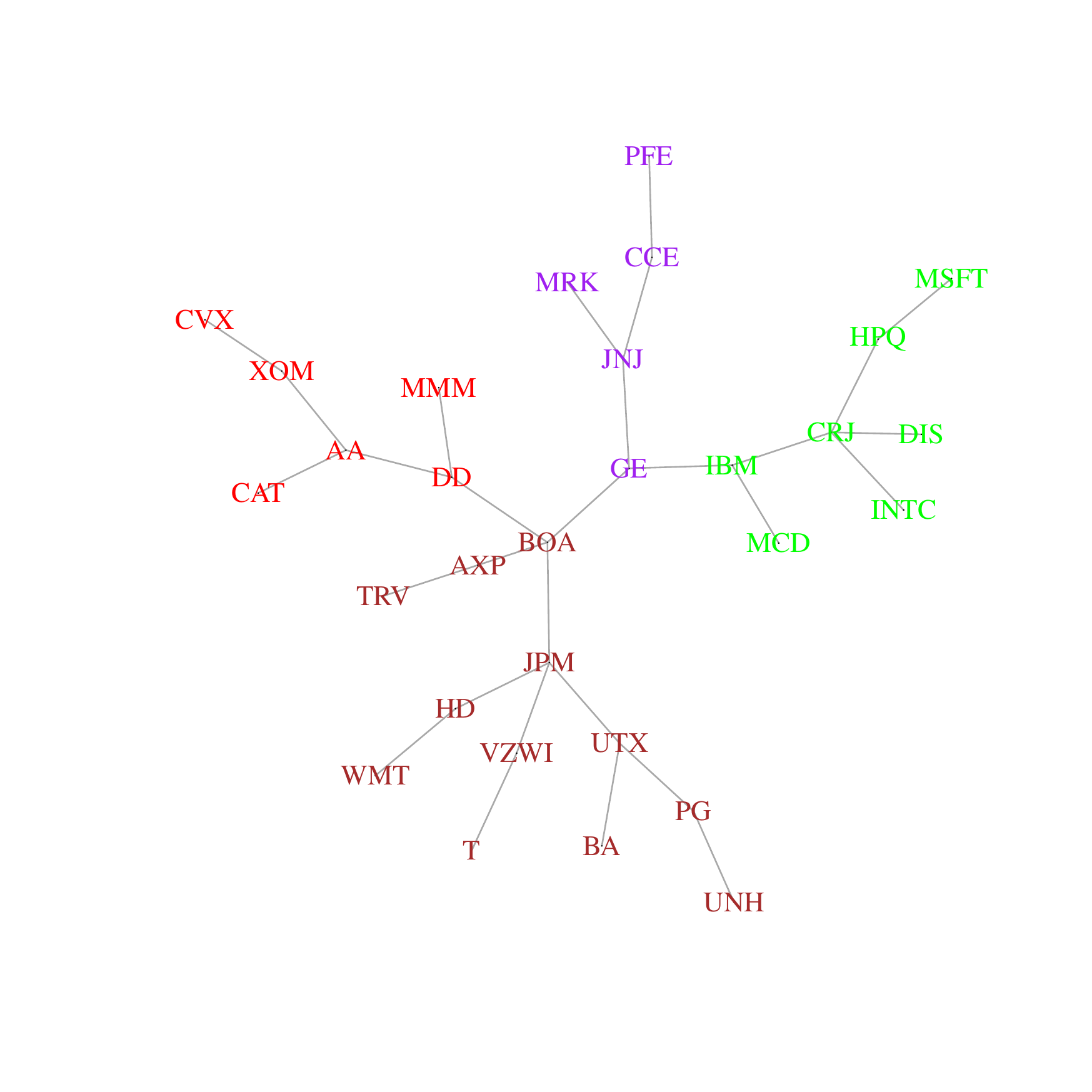}
  \caption{MST Period two DJIA with four clusters.}
  \label{fig:DJIAMST4CColoured}
\end{figure}

The task of dividing the MST into clusters can be difficult if
the MST's structure does not easily lend itself to being split into
the desired number of clusters. In Figure (\ref{fig:DJIAMST4CColoured}) 
we have the same MST with four clusters identified. It would be more
natural to only have three clusters with the branches rooted at
GE and IBM combined into a single cluster. 

The stock BOA is central within the MST so there is a question of which
cluster should it be assigned to, and, also which cluster should the
two stock branch of AXP and TRV be assigned to. Clearly AXP and TRV must
go to whatever cluster BOA is assigned. In Table (\ref{tab:BOAAXPDistances})
we have the distances between BOA and AXP and each of their
four nearest neighbors. The first column tells us that BOA should be
assigned to the brown cluster. The second column also shows that the
three nearest neighbors to AXP are in the brown cluster hence taking
it with BOA was reasonable.

Assigning cluster groups to pairs was relatively straight-forward for the
MSTs. In Figure (\ref{fig:DJIAMST4CColoured}) the green cluster is
opposite the red cluster so are paired and the remaining two 
lie opposite each other and are paired as well.
Further MSTs are in Appendices \ref{app:Period1Graphs} and
\ref{app:Period3Graphs}.

\subsection{Neighbor-Net Splits Graph}

\begin{figure}[ht]
  \centering
  \includegraphics[width=12cm]{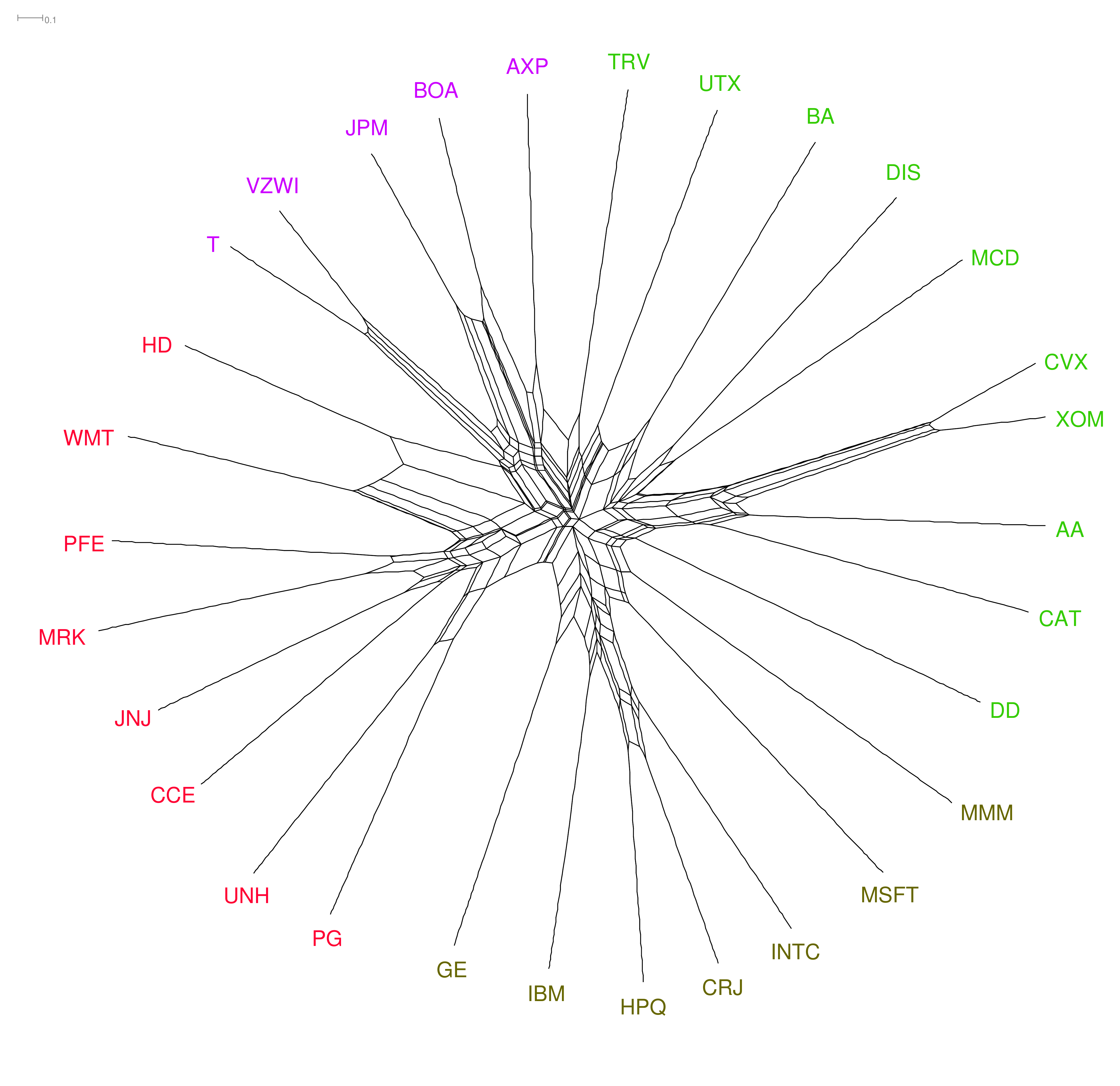}
  \caption{Neighbor-Net splits graph for  Period 2 DJIA with four clusters.}
  \label{fig:DJIANN4CColoured}
\end{figure}

In Figure (\ref{fig:DJIANN4CColoured}) shows the neighbor-Nets
splits graph for period two with four clusters identified. When 
defining correlation cluster the neighbor-Nets splits graphs have
a distinct advantage over either an HCT or an MST because the 
taxa, here stocks, are given a position in a circular ordering. The network
structure has some fairly clear breaks in its structure which allow
us to assign the stocks to clusters. Perhaps the clearest is the
red cluster; there is a clear break between GE and PG and again between
HD and T. Sometimes it is ambiguous which cluster a stock should be
assigned to, for example, DD could easily have been assigned to the 
khaki cluster rather than green cluster. Because of the circular
ordering its two nearest neighbors are MMM and CAT and probably no
harm would have been done had the assignment of this stock been done
differently. It is also possible to examine the distance matrix as an aid
in the decision of which cluster to assign an isolated stock to, such as
DD. However, this may not be an easy task because of the way neighbor-Nets
generates the network.

Table (\ref{tab:DDDistances}) presents the five nearest stocks to DD.
The closest, AA, is indeed in the green cluster, but the second, fourth,
and fifth
closest, BOA, AXP, and JPM are
in the purple on the opposite side of the network. The third closest is in the
khaki. The examination of nearest neighbors is not as helpful as it was
in the MST case, nevertheless the assignment to the green cluster seems
reasonable.

\begin{table}
\begin{center}
\begin{tabular}{lr}
Stock & Distance \\
\hline
AA  & 1.035 \\
BOA & 1.047 \\
GE  & 1.067 \\
AXP & 1.070 \\
JPM & 1.071 \\
\hline
\end{tabular}
\end{center}
\caption{Table of Distances to decide which cluster to put DD into for 
Period 2.}\label{tab:DDDistances}
\end{table}

While Figure (\ref{fig:DJIANN4CColoured}) shows four clusters
it is fairly
easy to see how these could be combined into two clusters, the red and
purple would be combined as would the green and khaki alternatively the
red and khaki can be combined as would the green and purple. 

The circular ordering of the stocks makes the pairing of the clusters
straight forward. The khaki group would be paired with the purple and
the green with the red. Further NN splits graphs 
are in Appendices \ref{app:Period1Graphs} and
\ref{app:Period3Graphs}.

\subsection{Comparison of Clusters}
\begin{table}
\begin{tabular}{cccccccc}
Cluster & All & HCT-MST & HCT-NN & MST-NN & HCT & MST & NN  \\
\hline
1       & PG  &         & CCE    & AXP    &     &  AA  & \\
        & UNH &         & JNJ    & BOA    &     &  BA  &  \\
        &     &         & MRK    & HD     &     & CAT  & \\
        &     &         & PFE    & JPM    &     & CVX  & \\
        &     &         &        & T      &     & DD   & \\
        &     &         &        & VZWI   &     & MMM  &  \\
        &     &         &        & WMT    &     & TRV  & \\
        &     &         &        &        &     & UTX  & \\
        &     &         &        &        &     & XOM  & \\
\hline
2       & CRJ &         & AA     &        & AXP & CCE  & \\
        & DIS &         & BA     &        & BOA & JNJ  & \\
        & GE  &         & CAT    &        & HD  & MRK  & \\
        & HPQ &         & CXV    &        & JPM & PFE  & \\
        & IBM &         & DD     &        & T   &      & \\
        & INTC&         & MMM    &        & VZWI&      & \\
        & MCD &         & TRV    &        & WMT &      &  \\
        & MSFT&         & UTX    &        &     &      &  \\
        &     &         & XOM    &        &     &      &  \\
\hline
\end{tabular}
\caption{The distribution of stocks during period two across two clusters
for the three different methods of cluster identification. The column
``All'' is the stocks that are have been assigned to the same cluster by
all three methods. The next three columns are the stocks which
assigned to the same cluster by two of the methods, the remaining
three columns are the stocks that are unique to each method.}
\label{tab:p4CL}
\end{table}

Table (\ref{tab:p4CL}) shows the assignment of the stocks to two
clusters, see also Figures (\ref{fig:DJIAHCTP42ClColoured}), 
(\ref{fig:DJIAMSTP42ClColoured}) and (\ref{fig:DJIANN4CColoured}).
While the assignment of the labels 1 and 2 are arbitrary it is 
clear from Table (\ref{tab:p4CL}) that each method has both stocks
in common and stocks which are
different from the other two methods. A similar
analysis can be applied to the four clusters. The question
then is -- does this make any difference in portfolio selection
for very small, private investor sized, portfolios?

\section{Results}\label{sec:results}

\begin{table}
\begin{tabular}{lllllll}
Period 2           &        &              &     &         &  & \\
                   &        &              &     &         & Industry & Levene\\
Simulation results & Random & N-Net        & HCT & MST     &  Group & p-value \\
\hline
Mean return \\
(2-stock portfolios) & 60.80  & 62.77     & 76.49 &  64.77 & 65.84  \\
(4-stock portfolios) & 62.50  & 61.96     & 76.61 &  63.31 & 67.17 \\
(8-stock portfolios) & 62.31  & 62.66     & 76.33 &  62.54 & 66.32 \\
\hline
Standard Deviation \\
(2-stock portfolios) & 30.57 & 28.10      & 32.97 & 32.89 & 32.99 & $10^{-16}$\\
(4-stock portfolios) & 21.90 & 19.41      & 22.02 & 21.21 & 21.35 & 0.006\\
(8-stock portfolios) & 14.19 & 13.08      & 11.95 & 14.03 & 13.72 & 0.109\\
\hline
Sharpe Ratios \\
(2-stock portfolios) & 1.92 & 2.16       & 2.25*  & 1.90 & 1.93  \\
(4-stock portfolios) & 2.75 & 3.08       & 3.38*  & 2.88 & 3.04  \\
(8-stock portfolios) & 4.24 & 4.62       & 6.20*  & 4.30 & 4.67  \\
\hline 
\\
Period 3           &        &              &     &         & \\
                   &        &              &     &         & Industry& Levene \\
Simulation results & Random & N-Net        & HCT & MST     &  Group  &p-value\\
\hline
Mean return \\
(2-stock portfolios) & 24.72  & 22.09     & 35.48 &  24.46 & 24.52  \\
(4-stock portfolios) & 24.41  & 22.65     & 35.97 &  24.60 & 25.28 \\
(8-stock portfolios) & 25.40  & 21.93     & 36.49 &  25.09 & 25.29 \\
\hline
Standard Deviation \\
(2-stock portfolios) & 23.52 & 24.52      & 18.93 & 23.54 & 22.55 & 0.012 \\
(4-stock portfolios) & 16.07 & 16.20      & 12.35 & 15.83 & 14.65 & 0.042 \\
(8-stock portfolios) &\ 9.88 & 10.86      &\ 7.39 & 10.50 &  9.40 & 0.029\\
\hline
Sharpe Ratios \\
(2-stock portfolios) & 0.86 & 0.72       & 1.64*  & 0.85 & 0.89  \\
(4-stock portfolios) & 1.25 & 1.13       & 2.56*  & 1.28 & 1.43  \\
(8-stock portfolios) & 2.13 & 1.61       & 4.34*  & 1.97 & 2.22  \\
\hline
\end{tabular}
\caption{Mean portfolio returns and the standard deviations of 1000 replications
of five different portfolio selection methods for in-sample (period two) and
out-of-sample (period three). The Sharpe ratios used
a period return of 2.2\% and 4.4\% for periods two and three respectively. 
The highest Sharpe ratios within each
portfolio size are marked with an
asterisk. The Levene test p-values exclude the HCT because all results
which included the HCT were very highly significant.}\label{tab:period2-3}
\end{table}


\begin{table}
\begin{tabular}{llrrrrrrrr}
Period & Cluster & N-Net &       & HCT   &         & MST   & & Industry       \\
2      & 1       & 74.50 & (8)    & 76.0 & (2)      & 78.9 &  (9)  & 69.6 & (8) \\
In     & 2       & 63.00 & (5)    & 38.5 & (4)      & 59.0 &  (8)  & 102.6 & (5)   \\
Sample & 3       & 41.90 & (10)   & 145.0 & (2)      & 35.6 &  (7)  & 49.0 & (8) \\
       & 4       & 73.44 & (7)    & 59.1 & (22)     & 77.8 &  (6)  & 48.2 & (9) \\
\hline
3      & 1       & 26.75 & (8)    & -5.00 & (2)     & 24.1 &  (9)  & 7.9 & (8)  \\
Out of & 2       & 15.60 & (5)    &  43.74 & (4)    & 13.6 &  (8)  & 20.6 & (5)  \\
Sample & 3       & 27.30 & (10)   &  43.50 & (2)    & 33.6 &  (7)  & 30.6 & (10) \\
       & 4       & 24.14 & (7)    &  21.91 & (22)   & 28.8 &  (6)  & 37.4 & (7) \\
\hline
3      & 1 & 27.50 & (6)         & 56.5 & (2)      & 17.2 &  (5)  & 7.9 & (8) \\
In     & 2 & 29.27 & (11)        & 20.73 & (22)    & 29.8 &  (5)  & 20.6 & (5) \\
Sample & 3 & -4.14 & (7)         & -40.00 & (1)    & 34.8 &  (13) & 30.6 & (10) \\
       & 4 & 46.0 & (6)         &  41.00 & (5)    & 6.6  &  (7)  & 37.4 & (7) \\
\hline
4      & 1 & 122.0 & (6)          & 234.0 & (2)     & 119.2 &  (5) & 101.9 & (8) \\
Out of & 2 & 64.45 & (11)        & 89.90 & (22)    & 38.4  & (5)  & 77.4 & (5) \\
Sample & 3 & 84.28 & (7)         & 127.0 & (1)     & 139.3 &  (13) & 131.5 & (10) \\
       & 4 & 191.0 & (6)         & 121.0 & (5)     & 82.6  & (7)   & 94.3 & (7) \\
\end{tabular}
\caption{The per-period returns and cluster sizes for the three network
methods and the industry groups.}\label{tab:returnsperperiod}
\end{table}

In this section we discuss the results in Table (\ref{tab:period2-3})
which gives the results of the in-sample and out-of-sample portfolio
simulations for period two clusters for periods two and three. Results for other
periods are in Appendix \ref{sec:extraresults}.

In all cases the random selection method, as expected, 
gave mean returns which were
statistically indistinguishable from the mean return of the thirty
stocks. The random selection method is the only one of the five
in which all stocks are equally likely to be chosen for a simulated
portfolio. In the other four methods stocks in small clusters are
more likely to be included in a simulated portfolio than stocks
in large clusters. Thus the mean return for a simulated
portfolio depends on the
mean returns of the clusters. 

Table (\ref{tab:returnsperperiod}) presents some of these values of
the returns for periods two and three in-sample and periods three and
four out-of-sample. The exceptionally high returns of the HCT  seen
in period  two in Table (\ref{tab:period2-3}) are due to the fact that
one of the two-stock clusters had a mean return of 145\%. This is an
unexpected result because the returns were not used directly to
generated the HCTs. Instead, the returns were used to generate the correlations
and it is on the basis of the correlations that the HCTs were generated.
Thus, we have no reason to expect that the portfolios selected by
the HCTs would have higher than average returns. Yet that is what
we see both for in-sample and out-of-sample testing for all periods.

In all but two cases the Levene test of equality of variances were
 statistically significant. The results in Tables (\ref{tab:period2-3}),
(\ref{tab:period1}), and (\ref{tab:period6}) exclude the HCT from the
Levene test because we were concerned that the highly unbalanced cluster
sizes may depress the standard deviations and make the results look
statistically significant when, in fact, they were not. 

In the in-sample testing neighbor-Net portfolios all had lower
standard deviation compared to random selection, though these were
only statistically significant in seven of the nine cases.
Figures for the HCT, MST and Industry selection methods were
seven, six and eight out of nine respectively.
This shows that each of these methods was  extracting 
useful financial
information from the data but, with the possible exception of
 neighbor-Net, was no better than dividing stocks into groups
by industry. 

Unfortunately this did not carry over
to the out-of-sample tests where 
the reduction in standard deviation occurred four times for neighbor-Net,
eight times for HCT and  once for MST out of nine sets of
simulations. 

For the industry group selection
there is no difference between in-sample and out-of-sample tests 
because the industry groups were constant across all simulations.
They showed a reduction variance with respect to random selection
in eight out of 12 sets of simulations.

\section{Discussion and Conclusions}\label{sec:discussion}

The results of the simulated portfolios show that the portfolios
selected using the correlation clusters identified by the HCTs outperformed 
all other portfolio selection methods based on the Sharpe ratio, with only
one exception.  This was, in part, because of high returns (the numerator
of the Sharpe ratio) relative to the other portfolio selection methods.
Other times a low standard deviation (the denominator) also
contributed to the high Sharpe ratio. The low standard
deviations are understandable given the way the simulations were run
and the highly unbalanced cluster sizes, but the high returns were 
unexpected. This needs to be checked with other markets.

Despite these excellent
results,
the highly unbalanced clusters 
sizes of the HCT clusters in all periods
calls into question whether HCTs should be used in
practice. For example, if one was choosing a four stock portfolio
based on period 2 clusters one would always include one of UNH and
PG and one of XOM and CVX, leaving the remaining two stocks to be
selected from 26. It is questionable whether an investor would accept
such a severe restriction in their stock selection.

Numerous papers have demonstrated the value of graph theory methods in
shedding light on the correlation structure of a stock market. This
paper in no way negates the value of those papers. Indeed,
the fact that the neighbor-Net portfolios all had lower
standard deviation compared to random selection in the in-sample
testing shows that neighbor-Net was extracting useful financial
information from the data. However, it is clear
that it is not straight forward to apply these insights in portfolio
selection. Thus,
a potential direction for future research is to explore
ways to combine the insights
provided by the graph theory methods with other information about the
stocks in order to improve stock selection.

The industry groups used here may not be particularly meaningful
because the original 21 industry sectors represented by the
30 stocks were combined into four
super-groups. In these circumstances one may argue that random selection
without replacement would make better use of the 21 industries represented
than the four super-groups we did use. Future research should use
a larger number of stocks in order to give a more meaningful
set of industry groups for selection in the portfolio simulations.

The type of reduction in portfolio standard deviation used in these simulations
is important because if an investor chose to hold a portfolio that was
under-diversified relative to the broad market a significant concern
would be that their portfolio would under-perform the market. A
low standard deviation means that in the case of under-performance
their actual performance would likely be closer to that of the market than
a method which had a high standard deviation. 

In our results the differences in standard deviations seem so small that
they are unlikely to be of economic significance. Consequently,
our results show that the
standard advice to hold a well-diversified portfolio, which in
practice means a larger portfolio than we have tested here, or better, buy
a low cost index fund, is good advice.

\bibliography{StockNNet}

\begin{thebibliography}{}

\bibitem[\protect\citeauthoryear{Bai and Green}{Bai and Green}{2010}]{Bai2010}
Bai, Y. and C.~J. Green (2010).
\newblock {International Diversification Strategies: Revisited from the Risk
  Perspective}.
\newblock {\em The Journal of Banking and Finance\/}~{\em 34}, 236--245.

\bibitem[\protect\citeauthoryear{Bali, Cakici, Yan, and Zhang}{Bali
  et~al.}{2005}]{Bali2005}
Bali, T.~G., N.~Cakici, X.~Yan, and Z.~Zhang (2005).
\newblock {Does Idiosyncratic Risk Really Matter?}
\newblock {\em The Journal of Finance\/}~{\em 60\/}(2), 905--929.

\bibitem[\protect\citeauthoryear{Barber and Odean}{Barber and
  Odean}{2008}]{Barber2008}
Barber, B.~M. and T.~Odean (2008).
\newblock {All That Glitters: The Effect of Attention and News on the Buying
  Behaviour of Individual and Institutional Investors}.
\newblock {\em The Review of Financial Studies\/}~{\em 21\/}(2), 785--818.

\bibitem[\protect\citeauthoryear{Benzoni, Collin-Dufresne, and
  Goldstein}{Benzoni et~al.}{2007}]{Benzoni2007}
Benzoni, L., P.~Collin-Dufresne, and R.~S. Goldstein (2007).
\newblock {Portfolio Choice over the Life-Cycle when the Stock and Labor
  Markets are Cointegrated}.
\newblock {\em The Journal of Finance\/}~{\em 62\/}(5), 2123--2167.

\bibitem[\protect\citeauthoryear{Bonanno, Calderelli, Lillo, Miccich\'{e},
  N.Vandewalle, and Mantegna}{Bonanno et~al.}{2004}]{Bonanno2004}
Bonanno, G., G.~Calderelli, F.~Lillo, S.~Miccich\'{e}, N.Vandewalle, and R.~N.
  Mantegna (2004).
\newblock {Networks of equities in financial markets}.
\newblock {\em The European Physical Journal B\/}~{\em 38}, 363--371.
\newblock doi:10.1140/epjb/e2--4-00129-6.

\bibitem[\protect\citeauthoryear{Bryant and Moulton}{Bryant and
  Moulton}{2004}]{Bryant2004}
Bryant, D. and V.~Moulton (2004).
\newblock Neighbor-net: An agglomerative method for the construction of
  phylogenetic networks.
\newblock {\em Molecular Biology and Evolution\/}~{\em 21\/}(2), 255--265.

\bibitem[\protect\citeauthoryear{Cont}{Cont}{2001}]{Cont2001}
Cont, R. (2001).
\newblock {Empirical properties of asset returns: stylized facts and
  statistical issues}.
\newblock {\em Quantitative Finance\/}~{\em 1:2}, 223--236.

\bibitem[\protect\citeauthoryear{Csardi and Nepusz}{Csardi and
  Nepusz}{2006}]{igraph}
Csardi, G. and T.~Nepusz (2006).
\newblock The igraph software package for complex network research.
\newblock {\em InterJournal\/}~{\em Complex Systems}, 1695.

\bibitem[\protect\citeauthoryear{DiMiguel, Garlappi, and Uppal}{DiMiguel
  et~al.}{2009}]{DeMiguel2009}
DiMiguel, V., L.~Garlappi, and R.~Uppal (2009).
\newblock {Optimal versus Naive Diversification: How Inefficient is the 1/N
  Portfolio Strategy?}
\newblock {\em Review of Financal Studies\/}~{\em 22\/}(5), 1915--1953.

\bibitem[\protect\citeauthoryear{Djauhari}{Djauhari}{2012}]{Djauhari2012}
Djauhari, M.~A. (2012).
\newblock {A Robust Filter in Stock Networks Analysis}.
\newblock {\em Physica A\/}~{\em 391\/}(20), 5049--5057.

\bibitem[\protect\citeauthoryear{Domian, Louton, and Racine}{Domian
  et~al.}{2007}]{Domian2007}
Domian, D.~L., D.~A. Louton, and M.~D. Racine (2007).
\newblock {Diversification in Portfolios of Individual Stocks: 100 Stocks Are
  Not Enough}.
\newblock {\em The Financial Review\/}~{\em 42}, 557--570.

\bibitem[\protect\citeauthoryear{Evans and Archer}{Evans and
  Archer}{1968}]{Evans1968}
Evans, J.~L. and S.~H. Archer (1968).
\newblock {Diversification and the Reduction of Dispersion: An Empirical
  Analysis}.
\newblock {\em The Journal of Finance\/}~{\em 23\/}(5), 761--767.

\bibitem[\protect\citeauthoryear{Fama and French}{Fama and
  French}{1992}]{Fama1992}
Fama, E.~F. and K.~R. French (1992).
\newblock {The Cross-Section of Expected Stock Returns}.
\newblock {\em The Journal of Finance\/}~{\em 47\/}(2), 427--465.

\bibitem[\protect\citeauthoryear{French and Fama}{French and
  Fama}{1993}]{Fama1993}
French, K.~R. and E.~F. Fama (1993).
\newblock {Common Risk Factors in the Returns on Stocks and Bonds}.
\newblock {\em Journal of Financal Economics\/}~{\em 33}, 3--56.

\bibitem[\protect\citeauthoryear{Gastwirth, Gel, Hui, Lyubchich, Miao, and
  Noguchi}{Gastwirth et~al.}{2013}]{lawstat}
Gastwirth, J.~L., Y.~R. Gel, W.~L.~W. Hui, V.~Lyubchich, W.~Miao, and
  K.~Noguchi (2013).
\newblock {\em lawstat: An R package for biostatistics, public policy, and
  law}.
\newblock R package version 2.4.1.

\bibitem[\protect\citeauthoryear{Goyal and Santa-Clara}{Goyal and
  Santa-Clara}{2003}]{Goyal2003}
Goyal, A. and P.~Santa-Clara (2003).
\newblock {Idiosyncratic Risk Matters!}
\newblock {\em The Journal of Finance\/}~{\em 58\/}(3), 975--1007.

\bibitem[\protect\citeauthoryear{Jorion}{Jorion}{1985}]{Jorion1985}
Jorion, P. (1985).
\newblock {International Portfolio Diversification with Estimation Risk}.
\newblock {\em The Journal of Business\/}~{\em 58\/}(3), 259--278.

\bibitem[\protect\citeauthoryear{Kenett, Tumminello, Madi, Gur-Gershgoren,
  Mantegna, and Ben-Jacob}{Kenett et~al.}{2010}]{Kennet2010}
Kenett, D.~Y., M.~Tumminello, A.~Madi, G.~Gur-Gershgoren, R.~N. Mantegna, and
  E.~Ben-Jacob (2010).
\newblock Systematic analysis of group identification in stock markets.
\newblock {\em PLoS ONE\/}~{\em 5}, e15032.

\bibitem[\protect\citeauthoryear{Lowenfeld}{Lowenfeld}{1909}]{lowenfeld1909}
Lowenfeld, H. (1909).
\newblock {\em {Investment, an Exact Science}}.
\newblock Financial Review of Reviews.

\bibitem[\protect\citeauthoryear{Mantegna}{Mantegna}{1999}]{Mantegna1999}
Mantegna, R.~N. (1999).
\newblock {Hierarchical structure in financial markets}.
\newblock {\em The European Physical Journal B\/}~{\em 11}, 193--197.

\bibitem[\protect\citeauthoryear{Markowitz}{Markowitz}{1991}]{markowitz1959}
Markowitz, H.~M. (1991).
\newblock {\em {Portfolio Selection: Efficient Diversification of Investments
  2nd Edition}}.
\newblock Wiley.

\bibitem[\protect\citeauthoryear{Markowtiz}{Markowtiz}{1952}]{markowitz1952}
Markowtiz, H. (1952).
\newblock {Portfolio Selection}.
\newblock {\em The Journal of Finance\/}~{\em 7\/}(1), 77--91.

\bibitem[\protect\citeauthoryear{Miccich\'{e}, Bonannon, Lillo, and
  Mantegna}{Miccich\'{e} et~al.}{2006}]{Micciche2006}
Miccich\'{e}, S., G.~Bonannon, F.~Lillo, and R.~N. Mantegna (2006).
\newblock {Degree stability of a minimum spanning tree of price return and
  volatility}.
\newblock {\em Physica A\/}~{\em 324}, 66--73.

\bibitem[\protect\citeauthoryear{Michaud}{Michaud}{1989}]{Michaud1989}
Michaud, R.~O. (1989).
\newblock {The Markowitz Optimisation Enigma: is `Optimized' Optimal?}
\newblock {\em Financial Analysts Journal\/}~{\em 45\/}(1), 31--42.

\bibitem[\protect\citeauthoryear{Naylor, Rose, and Moyle}{Naylor
  et~al.}{2007}]{Naylor2007}
Naylor, M.~J., L.~C. Rose, and B.~J. Moyle (2007).
\newblock {Topology of foreign exchange markets using hierarchical structure
  methods}.
\newblock {\em Physica A\/}~{\em 382}, 199--208.

\bibitem[\protect\citeauthoryear{Onnela, Chakraborti, Kaski, Kert\`{e}sz, and
  Kanto}{Onnela et~al.}{2003a}]{Onnela2003a}
Onnela, J.-P., A.~Chakraborti, K.~Kaski, J.~Kert\`{e}sz, and A.~Kanto (2003a).
\newblock {Asset Trees and Asset Graphs in Financial Markets}.
\newblock {\em Physica Scripta\/}~{\em T106}, 48--54.

\bibitem[\protect\citeauthoryear{Onnela, Chakraborti, Kaski, Kert\`{e}sz, and
  Kanto}{Onnela et~al.}{2003b}]{Onnela2003}
Onnela, J.-P., A.~Chakraborti, K.~Kaski, J.~Kert\`{e}sz, and A.~Kanto (2003b).
\newblock {Dynamics of market correlations: Taxonomy and portfolio analysis}.
\newblock {\em Physical Review E\/}~{\em 68}, 0561101--1 -- 056110--12.

\bibitem[\protect\citeauthoryear{{R Core Team}}{{R Core Team}}{2014}]{R}
{R Core Team} (2014).
\newblock {\em R: A Language and Environment for Statistical Computing}.
\newblock Vienna, Austria: R Foundation for Statistical Computing.

\bibitem[\protect\citeauthoryear{Rea and Rea}{Rea and Rea}{2014}]{Rea2014}
Rea, A. and W.~Rea (2014).
\newblock {Visualization of a stock market correlation matrix}.
\newblock {\em Physica A\/}~{\em 400}, 109--123.

\bibitem[\protect\citeauthoryear{Sharpe}{Sharpe}{1964}]{Sharpe1964}
Sharpe, W.~F. (1964).
\newblock {Capital Asset Prices: A Theory of Market Equilibrium Under
  Conditions of Risk}.
\newblock {\em The Journal of Finance\/}~{\em 19\/}(3), 13--37.

\end{thebibliography}

\newpage

\appendix

\section{Period One Graphs}\label{app:Period1Graphs}

\subsection{Period One Hierarchical Clustering Trees}

\begin{figure}[ht]
  \centering
  \includegraphics[width=12cm]{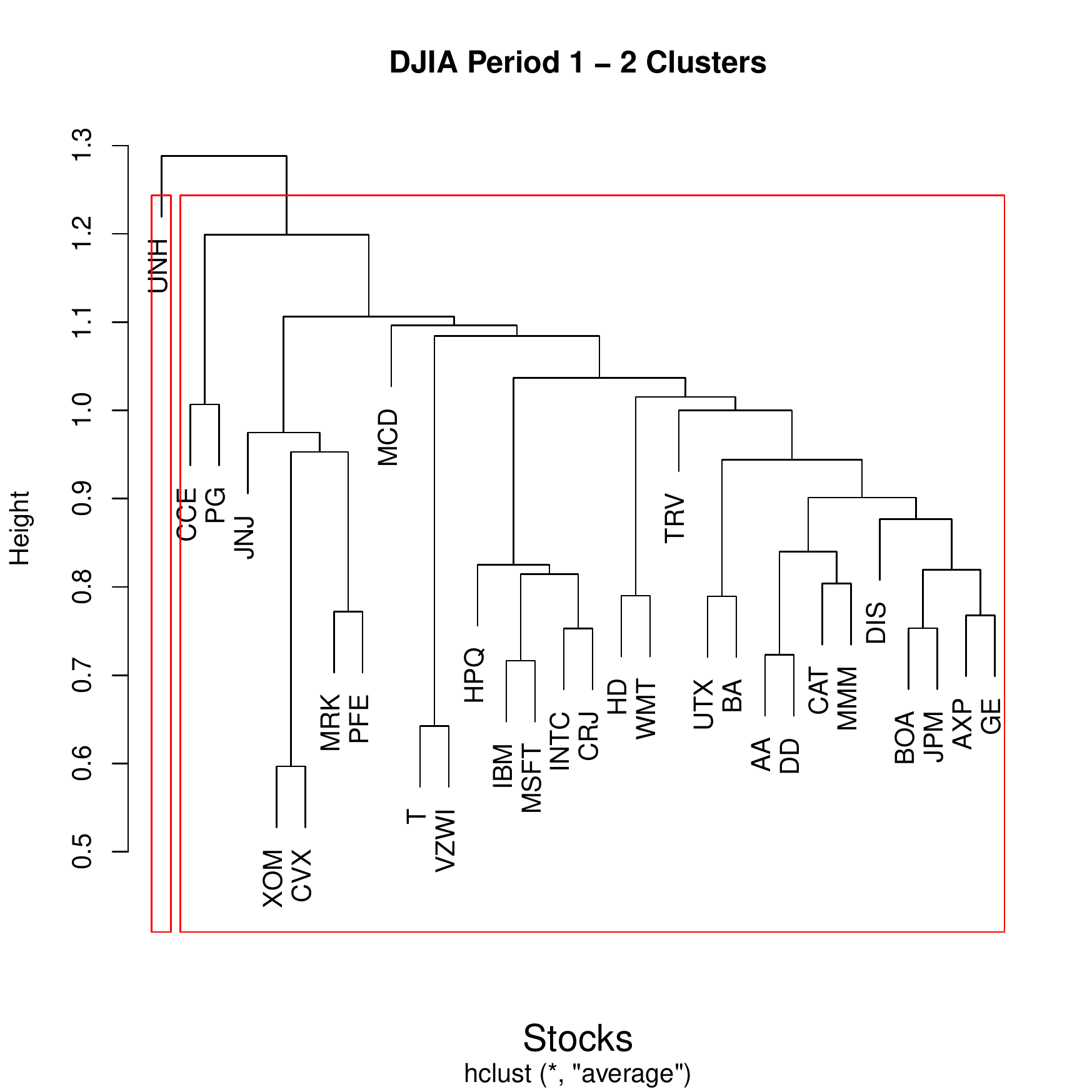}
  \caption{HCT Period 1 DJIA with two clusters. The unbalanced cluster
sizes are clearly seen with clusters of one and 29 stocks respectively.}
  \label{fig:DJIAHCTP1Cl2}
\end{figure}

\begin{figure}[ht]
  \centering
  \includegraphics[width=12cm]{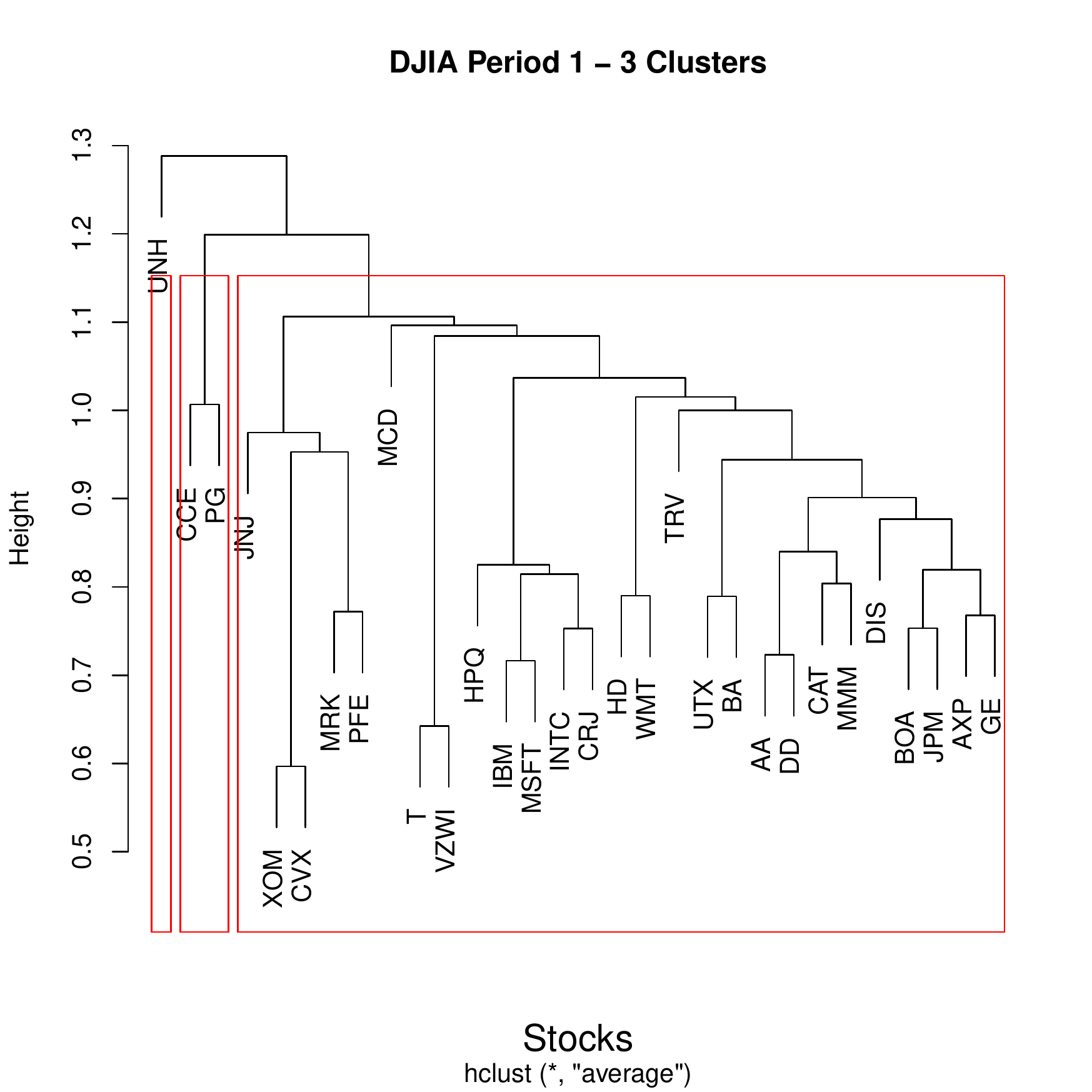}
  \caption{HCT Period 1 DJIA with three clusters. The unbalanced cluster
sizes are clearly seen with clusters of one, two  and 27 stocks respectively.}
  \label{fig:DJIAHCTP1Cl3}
\end{figure}

\begin{figure}[ht]
  \centering
  \includegraphics[width=12cm]{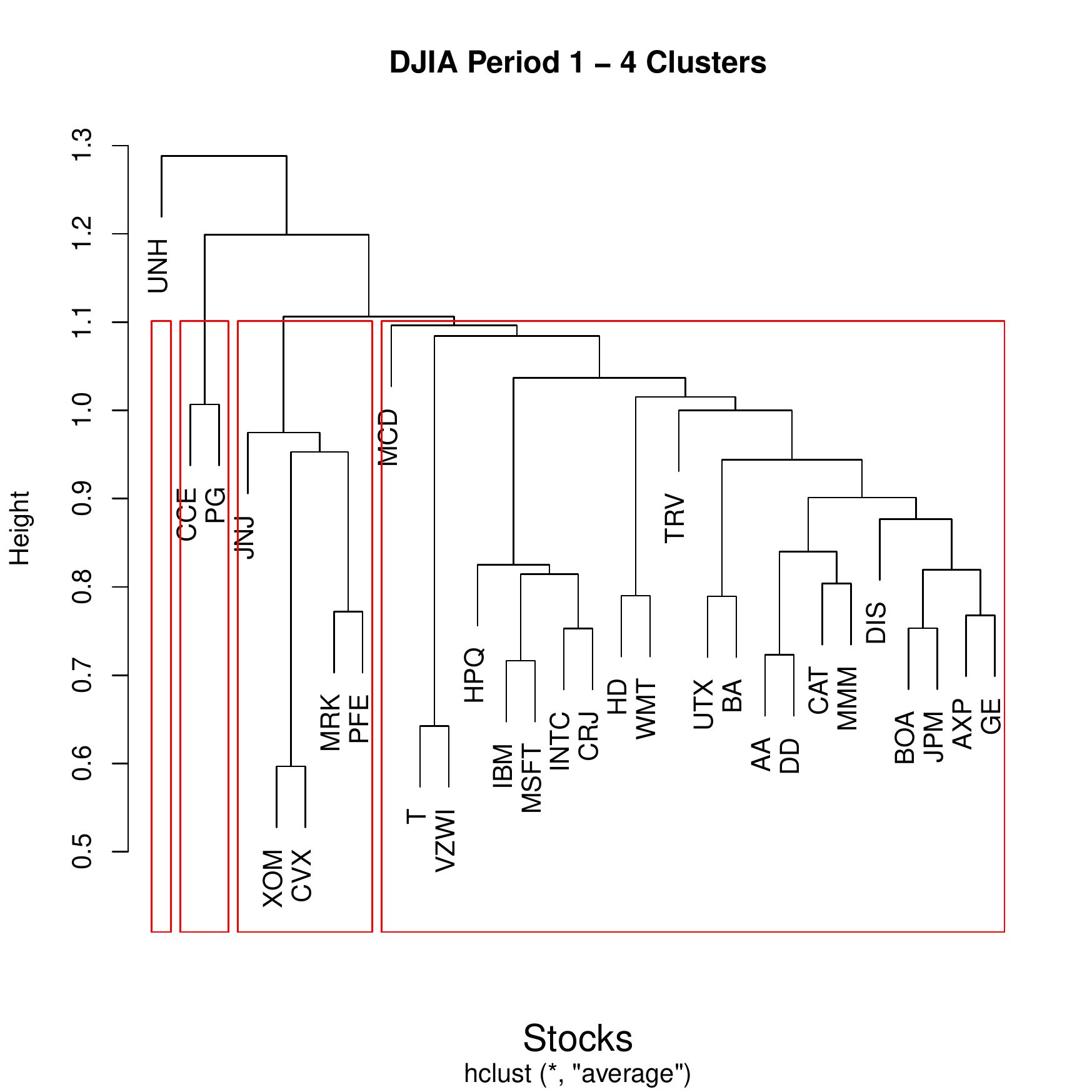}
  \caption{HCT Period 1 DJIA with four  clusters. The unbalanced cluster
sizes are clearly seen with clusters of one, two, five  and 22 stocks respectively.}
  \label{fig:DJIAHCTP1Cl4}
\end{figure}

\subsection{Period One Minimum Spanning Trees}

\begin{table}
\centering
\begin{tabular}{lr}
Stock & Distance \\
\hline 
INTC & 0.793 \\
MSFT & 0.804 \\
CRJ  & 0.850 \\
\hline
\end{tabular}
\caption{Table of the distances of the three stocks closest to HPQ
used to decide which cluster to assign HPQ to.}\label{tab:Period1HPQ}
\end{table}

\begin{figure}[ht]
  \centering
  \includegraphics[width=12cm]{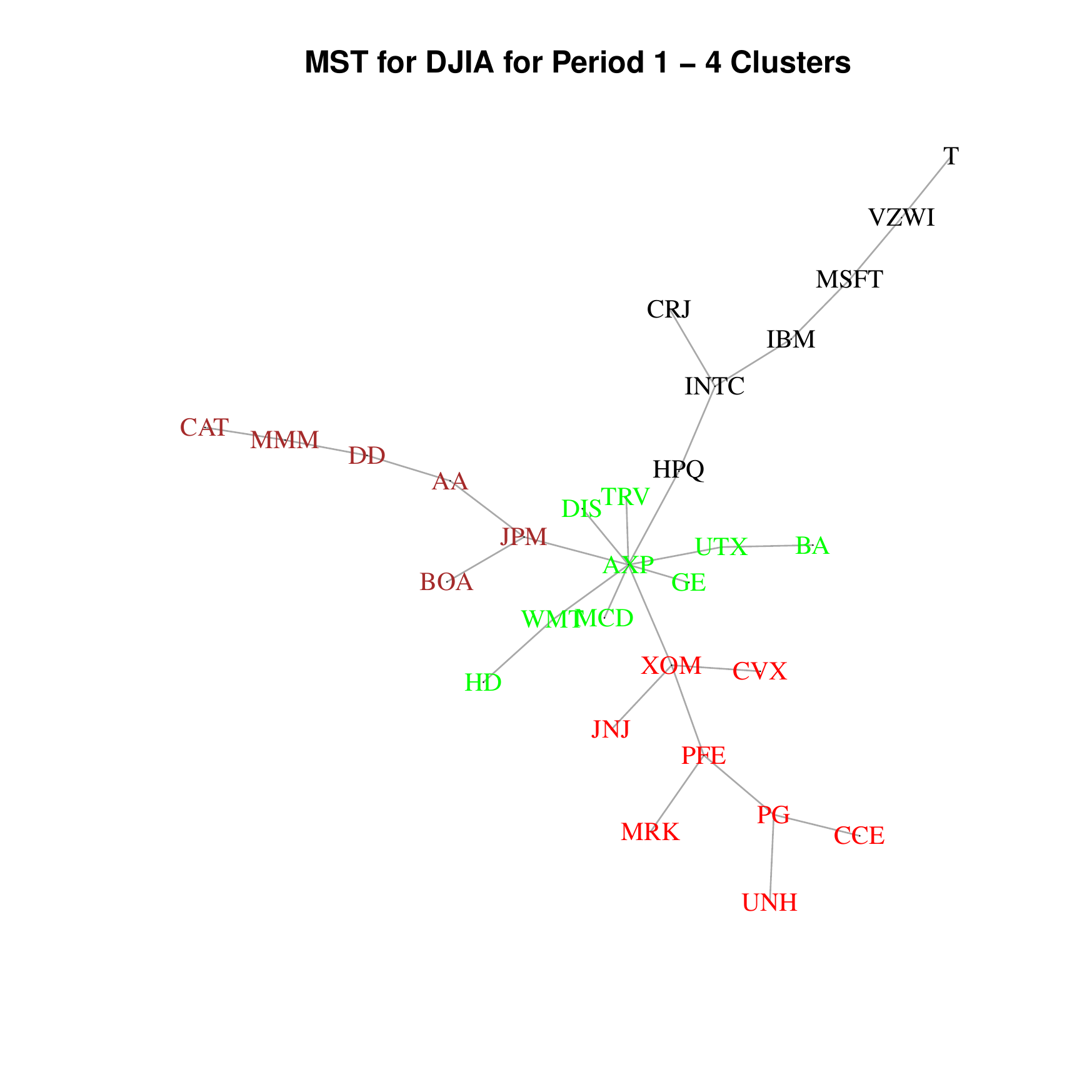}
  \caption{MST Period 1 DJIA with four  clusters. The cluster
sizes are well balanced with clusters of six, seven, eight  and nine 
stocks respectively.}
  \label{fig:DJIAMSTP1Cl4}
\end{figure}

The MST in Figure (\ref{fig:DJIAMSTP1Cl4}) shows AXP as the pivot stock
in the Dow 30. It has nine stocks to which is join as nearest neighbors.
Such a structure does not easily lend itself to dividing the 30 stocks into
groups. If we arbitrarily set the minimum branch size for a group to be
five stocks, there are three such branches in the MST. Of these branches
for the ones rooted at JPM and XOM it is clear which stocks are members of
these groups. For the branch rooted at HPQ there is a question of whether
HPQ should be assigned to the group containing AXP or INTC. Table
(\ref{tab:Period1HPQ}) shows the distances to the three nearest stocks
to HPQ. From this it is clear HPQ should be assigned to the group with
INTC.

\subsection{Period One Neighbor-Net Splits Graphs}

\begin{figure}[ht]
  \centering
  \includegraphics[width=12cm]{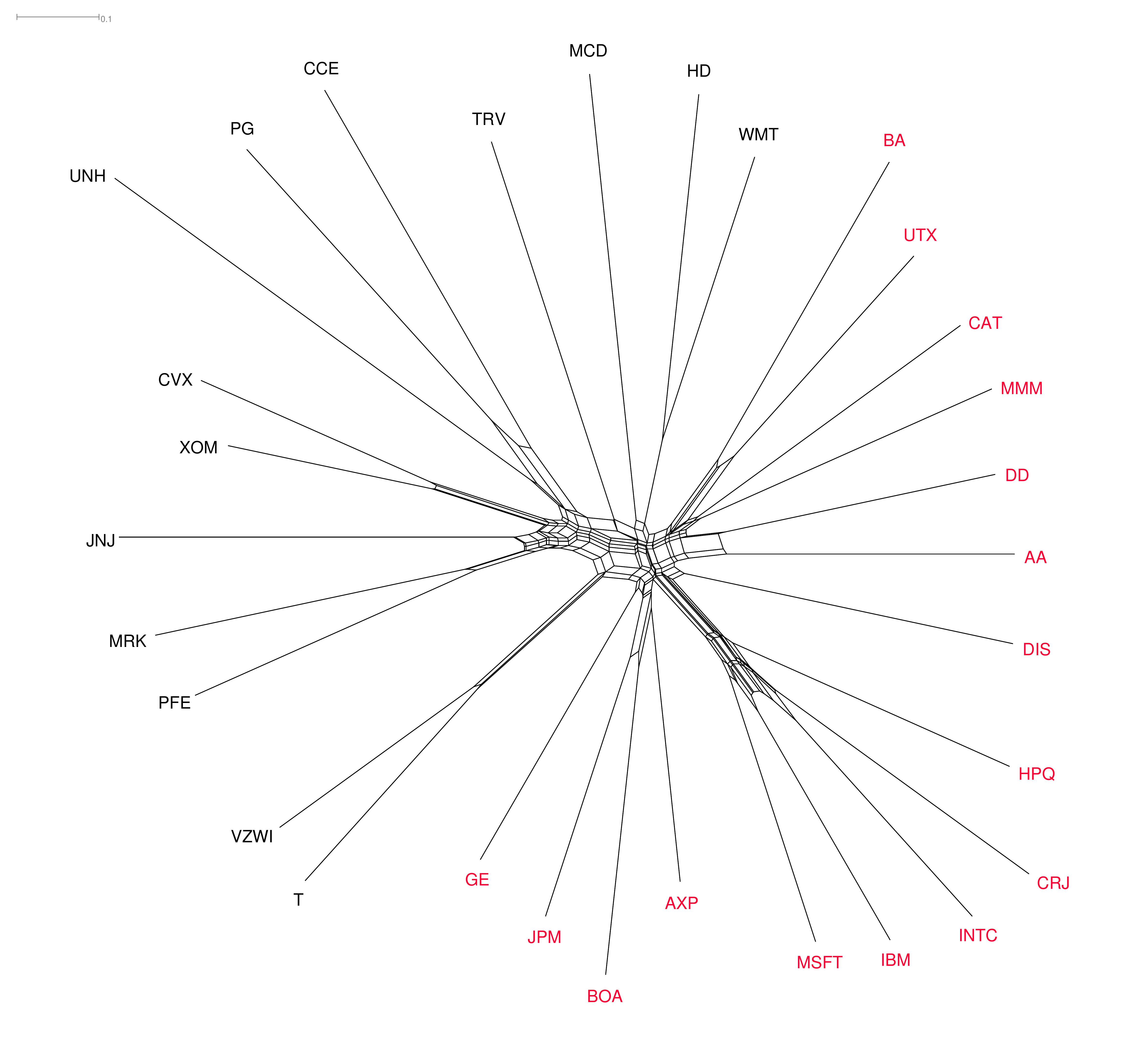}
  \caption{Neighbor-Net splits graph for
 Period 1 DJIA with two clusters. The cluster
sizes are well balanced with clusters of 14 and 16 
stocks respectively.}
  \label{fig:DJIANNP1Cl2}
\end{figure}

\begin{figure}[ht]
  \centering
  \includegraphics[width=12cm]{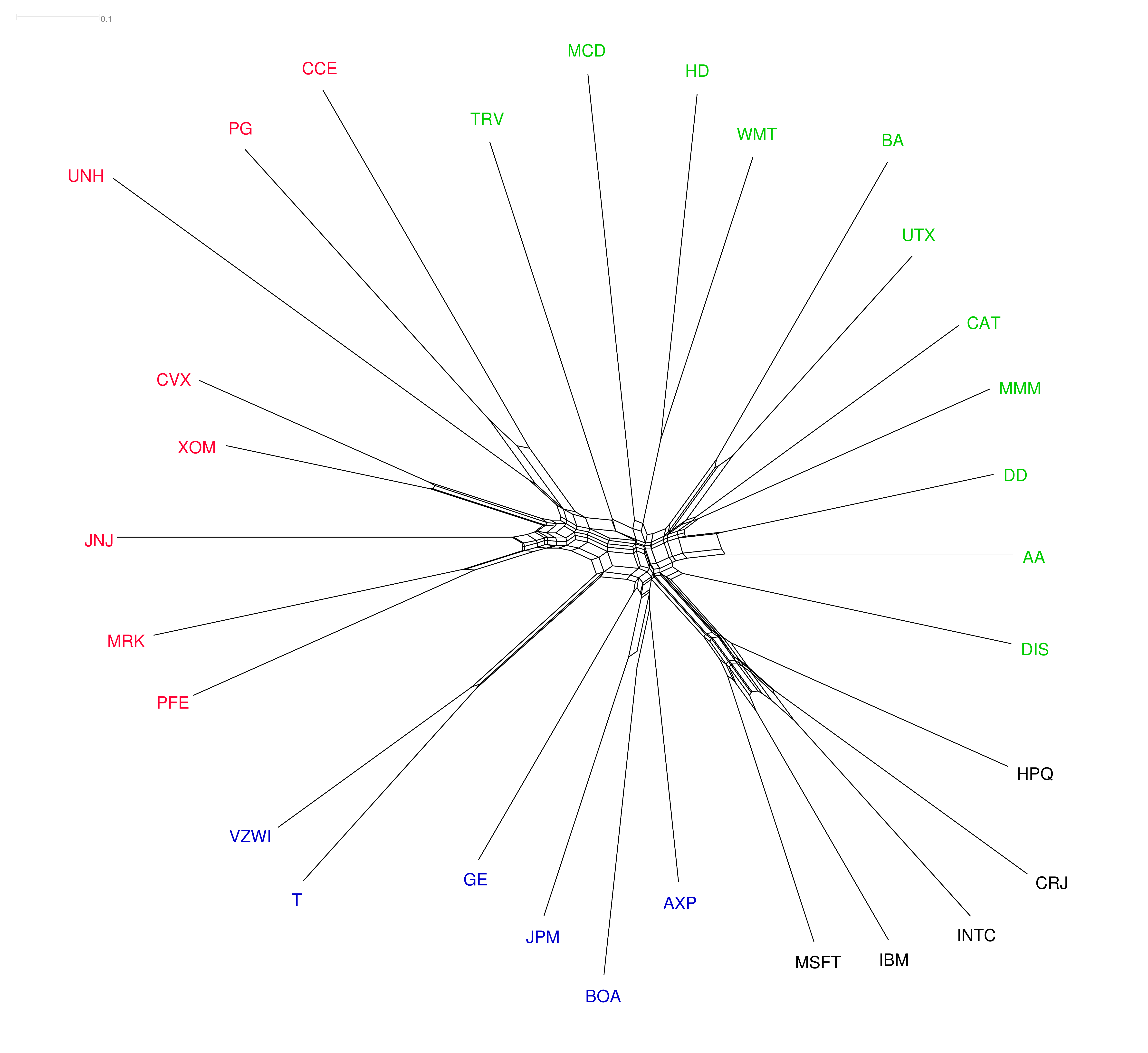}
  \caption{Neighbor-Net splits graph for
 Period 1 DJIA with four clusters. The cluster
sizes are somewhat unbalanced with clusters of five, six, nine and 11 
stocks respectively.}
  \label{fig:DJIANNP1Cl4}
\end{figure}

Figures (\ref{fig:DJIANNP1Cl2}) and (\ref{fig:DJIANNP1Cl4}) represent
the neighbor-Nets splits graphs for period one. The network structure has
as many as eight breaks allowing considerable flexibility in splitting
the stocks into groups. However, it is clear that the network structure
is elongated in the horizontal direction meaning that if the stocks
are to be split into two groups the most natural way to do that would be
to use a vertical cut. We have one way of doing that 
in Figure (\ref{fig:DJIANNP1Cl2}) where the
stocks have been split into two groups of size 14 and 16. 

In Figure (\ref{fig:DJIANNP1Cl4}) we have split the stocks into 
four groups. The network structure does not lend itself to easily
making four groups because there are isolated stocks such as TRV and MCD 
which are not close to any other stocks. Also there are some pairs such 
as VZMI and T 
on the lower side of the graphs and HD and WMT on the upper which
appear as isolated groups. In this case the circular ordering helps
because each of these isolated individual stocks or pairs of stocks
must be grouped with those either to the left or the right. In such a
case it can be helpful to look at a few of the stocks to see which
group is should be assigned to. For example, we consider TRV.

In Table (\ref{tab:NNTRVGroup}) it is clear that the three closest
stocks are in the group on the opposite side of the network while
the fourth and fifth closest are to the right. Thus TRV together
with MCD, HD, and WMT should all be included in the green rather than
the red group.

Had the distances in Table (\ref{tab:NNTRVGroup}) been taken into
account in the two group case (Figure \ref{fig:DJIANNP1Cl2})
it could have been argued that
TRV, MCD, HD, and WMT should be in the red group rather than the
black. Had this been done the clusters would be more unbalanced with
sizes of 11 and 19 stocks rather than the 14 and 16 we chose.

\begin{table}
\centering
\begin{tabular}{lr}
Stock & Distance \\
\hline
AXP & 0.783 \\
JPM & 0.859 \\
AA  & 0.866 \\
GE  & 0.870 \\
DD  & 0.874 \\
\hline
\end{tabular}
\caption{The five closest stocks to TRV in Period 1 and
their distance.}\label{tab:NNTRVGroup}
\end{table}

\clearpage

\section{Period Three Graphs}\label{app:Period3Graphs}


\subsection{Period Three Hierarchical Clustering Trees}

\begin{figure}[ht]
  \centering
  \includegraphics[width=12cm]{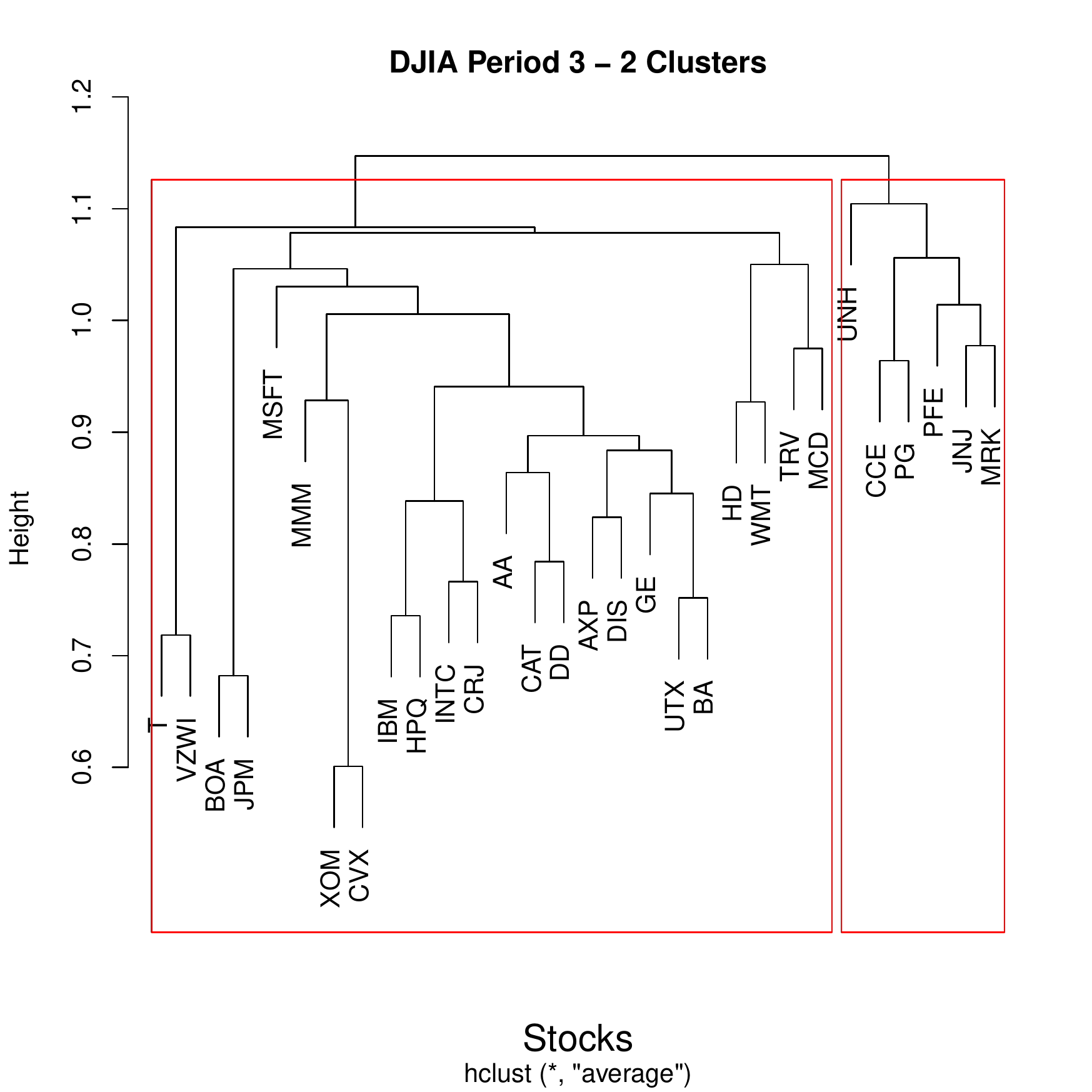}
  \caption{HCT Period 5 DJIA with two clusters. The unbalanced cluster
sizes are clearly seen with clusters of six and 24 stocks respectively.}
  \label{fig:DJIAHCTP52Cl2}
\end{figure}

\begin{figure}[ht]
  \centering
  \includegraphics[width=12cm]{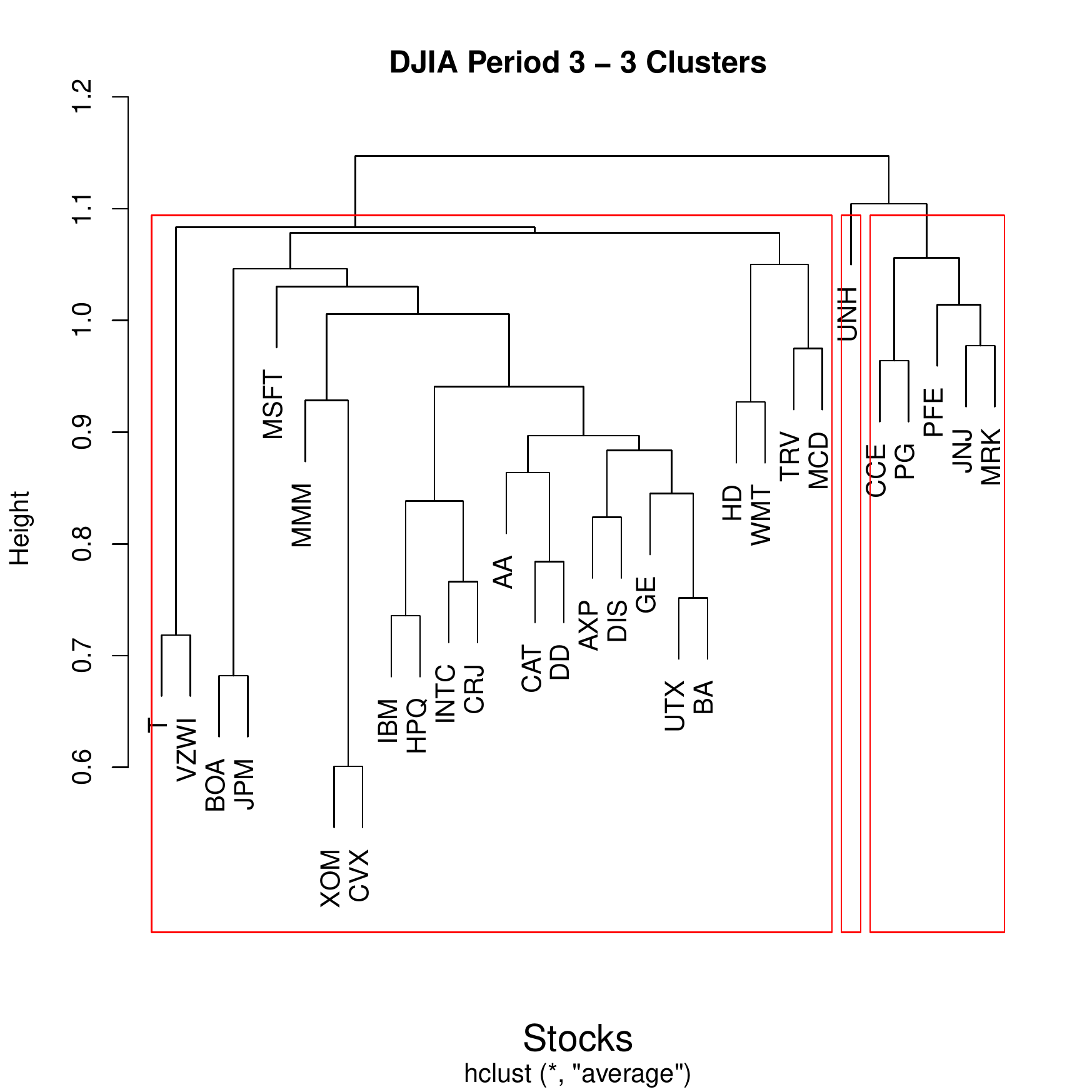}
  \caption{HCT Period 5 DJIA with three clusters. The unbalanced cluster
sizes are clearly seen with clusters of one, five and 24 stocks respectively.}
  \label{fig:DJIAHCTP52Cl3}
\end{figure}

\begin{figure}[ht]
  \centering
  \includegraphics[width=12cm]{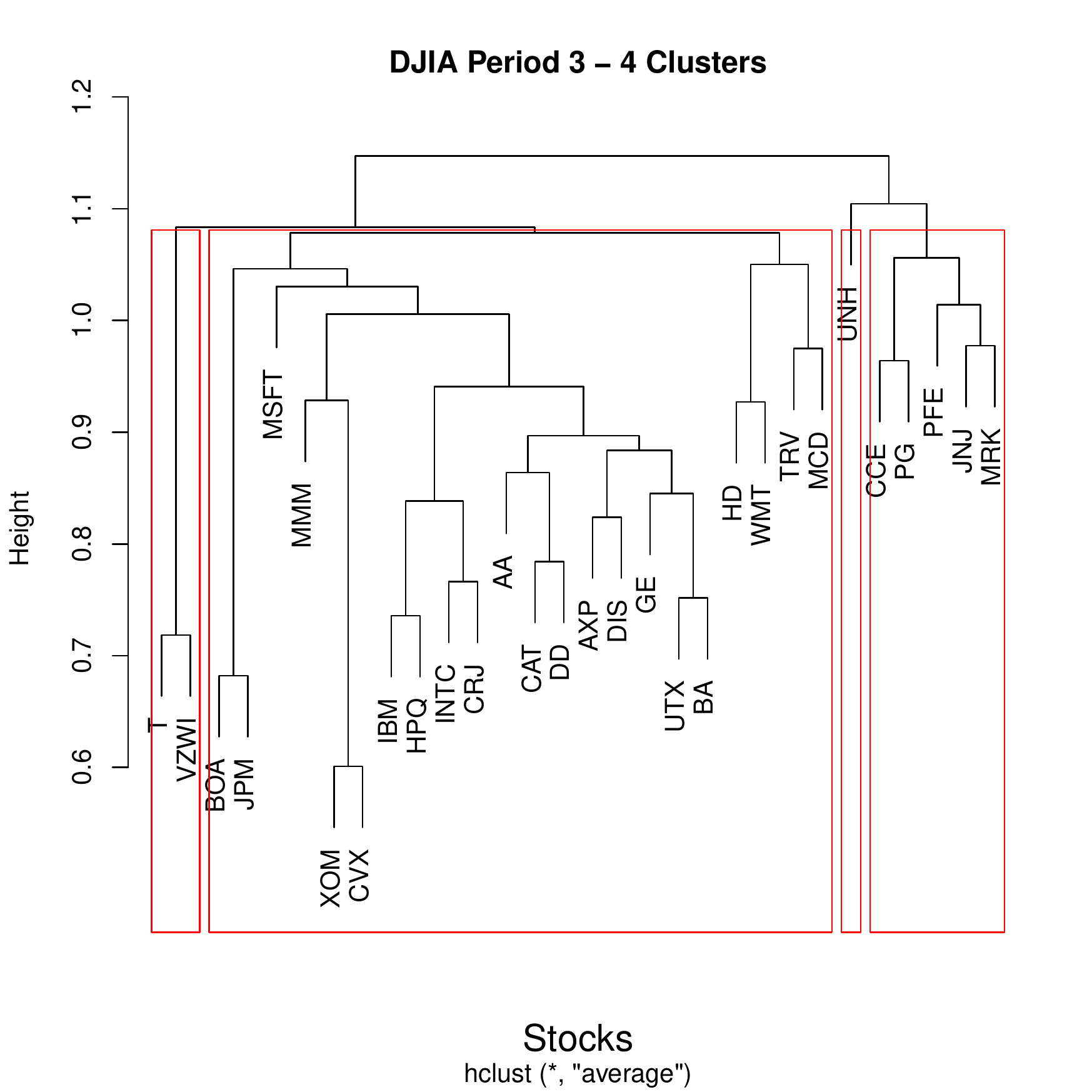}
  \caption{HCT Period 5 DJIA with four clusters. The unbalanced cluster
sizes are clearly seen with clusters of one, two, five and 22 stocks 
respectively.}
  \label{fig:DJIAHCTP52Cl4}
\end{figure}

\clearpage
\subsection{Period Three Minimum Spanning Trees}

\begin{figure}[ht]
  \centering
  \includegraphics[width=12cm]{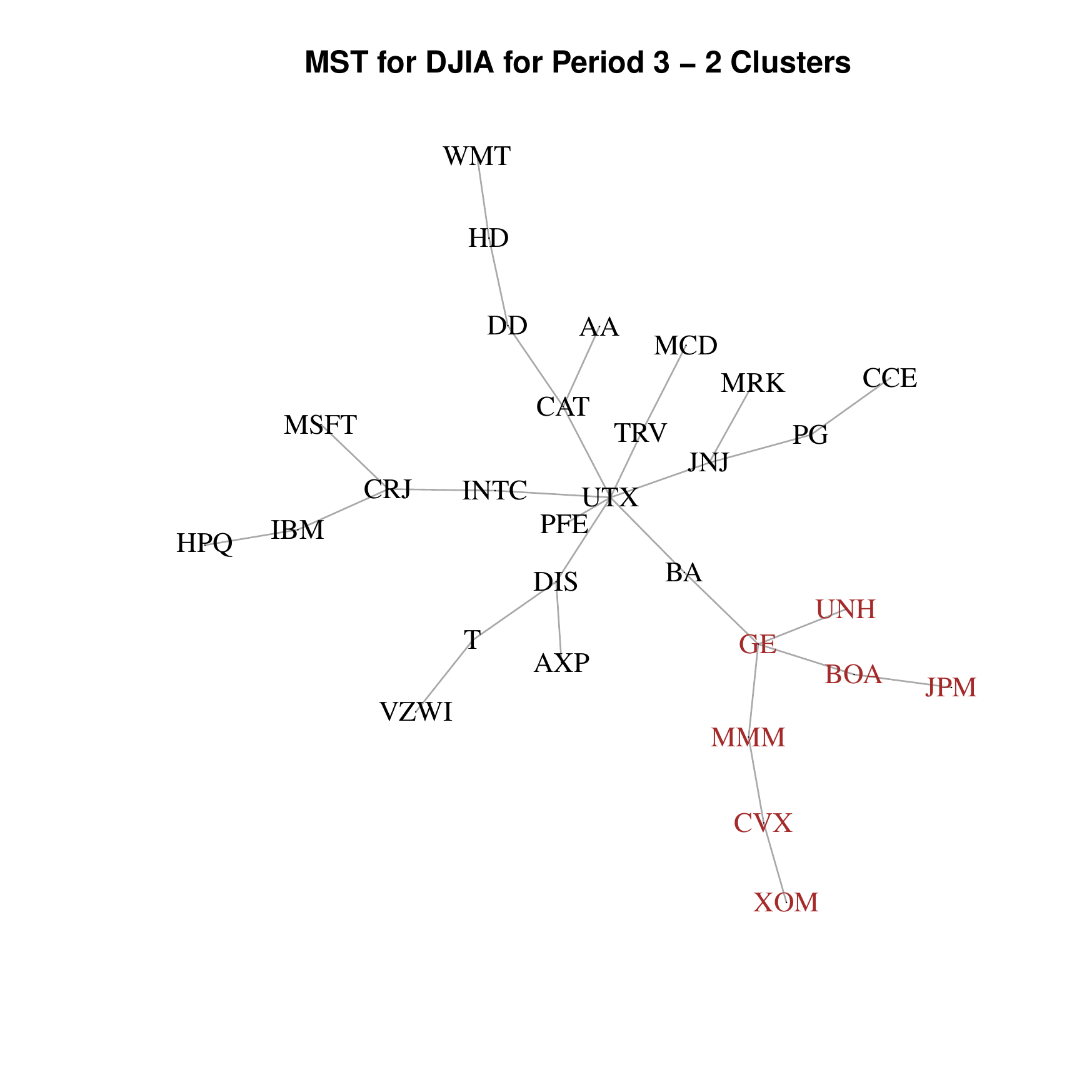}
  \caption{MST Period 3 DJIA with two clusters. The unbalanced cluster
sizes are clearly seen with clusters of seven and 23 stocks 
respectively.}
  \label{fig:DJIAMSTP5Cl2}
\end{figure}

\begin{figure}[ht]
  \centering
  \includegraphics[width=12cm]{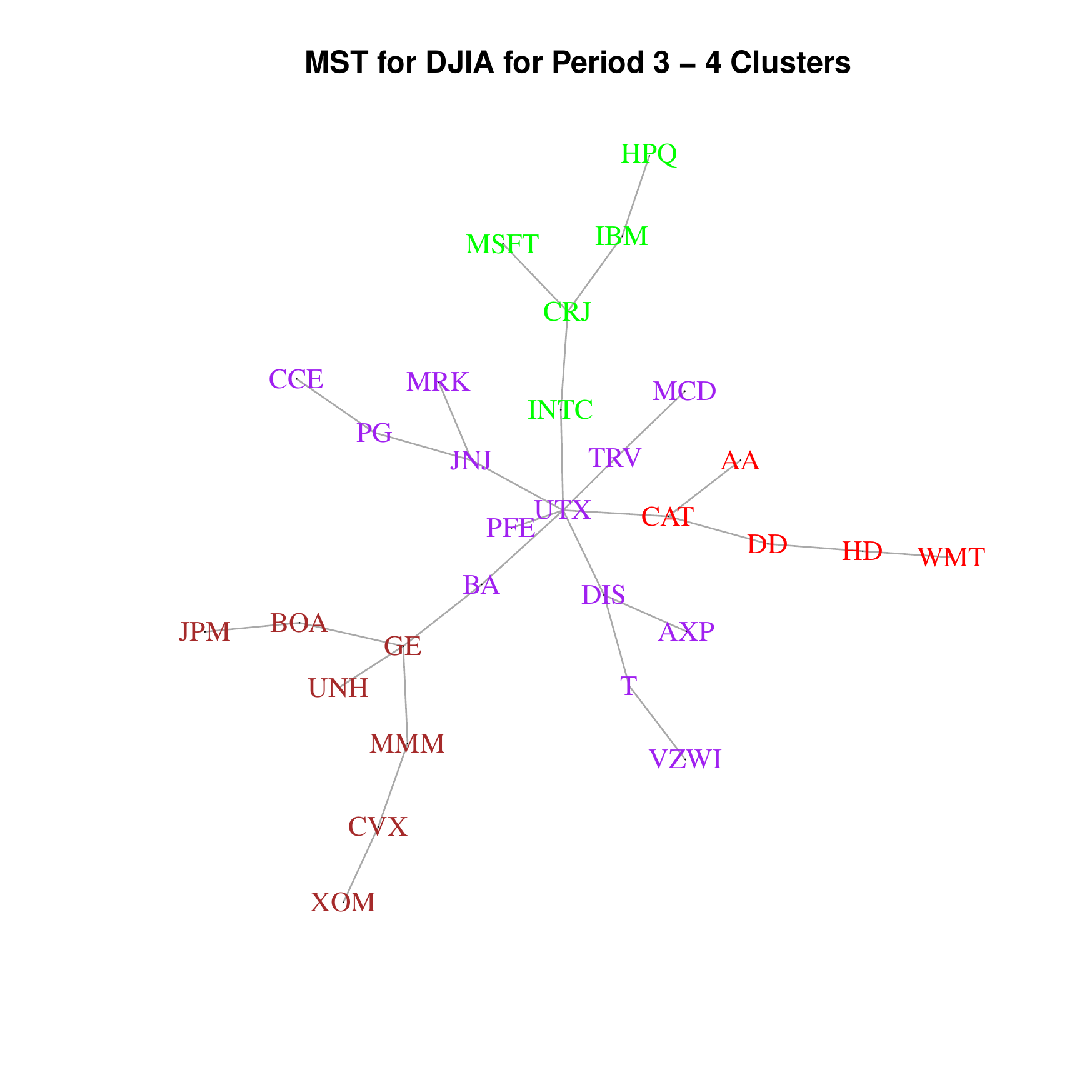}
  \caption{MST Period 3 DJIA with four clusters. The cluster
sizes are better balanced with clusters of five, five, seven and 13 stocks 
respectively.}
  \label{fig:DJIAMSTP5Cl4}
\end{figure}

In Figure (\ref{fig:DJIAMSTP5Cl2}) the pivotal nature of UTX and the short
branches extending from it give little flexibility in choosing two clusters.
The largest branch is connected to UTX via BA, so the question is whether
to assign BA to the larger or the smaller cluster.  Table 
(\ref{tab:BADistancesP5}) gives the nearest four stocks to BA. The first
two and fourth
closest stocks are in the large cluster. So while it would be nice to
assign BA to the smaller cluster in order to try to better balance the
cluster sizes, this clearly cannot be justified based on the data in
the table.

\begin{table}
\begin{center}
\begin{tabular}{lr}
Stock & Distance \\
\hline
UTX & 0.752 \\
DIS & 0.836 \\
GE  & 0.842 \\
AXP & 0.901 \\
\hline
\end{tabular}
\end{center}
\caption{Table of Distances to decide which cluster to put BA into for 
period 3.}\label{tab:BADistancesP5}
\end{table}

\clearpage

\subsection{Neighbor-Net Splits Graphs for Period Three}

\begin{figure}[ht]
  \centering
  \includegraphics[width=12cm]{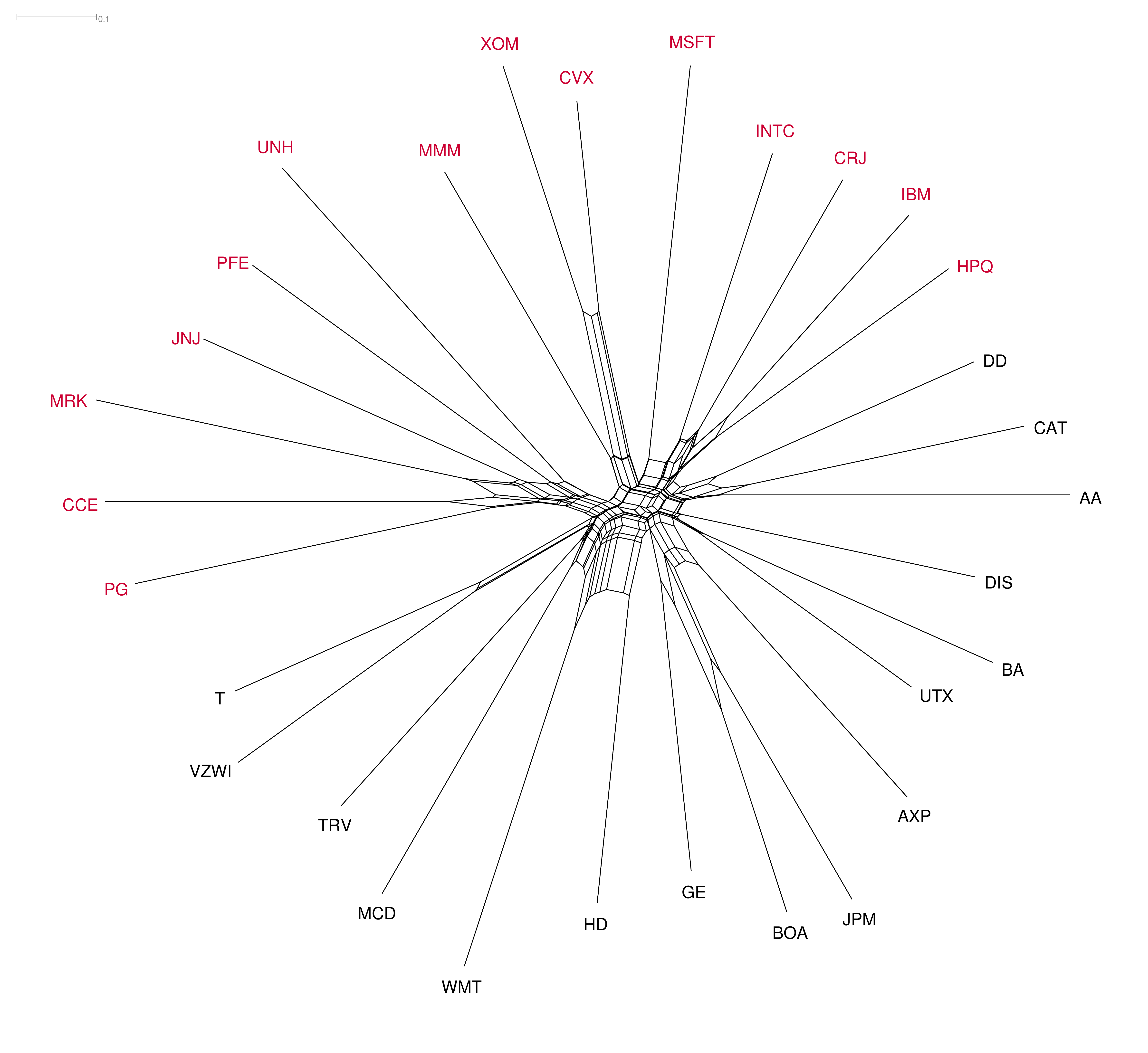}
  \caption{Neighbor-Net splits graph for  Period 3 DJIA with two clusters. 
The cluster
sizes are evenly balanced with clusters of 14 and 16 stocks 
respectively.}
  \label{fig:DJIANNP5Cl2}
\end{figure}

\begin{figure}[ht]
  \centering
  \includegraphics[width=12cm]{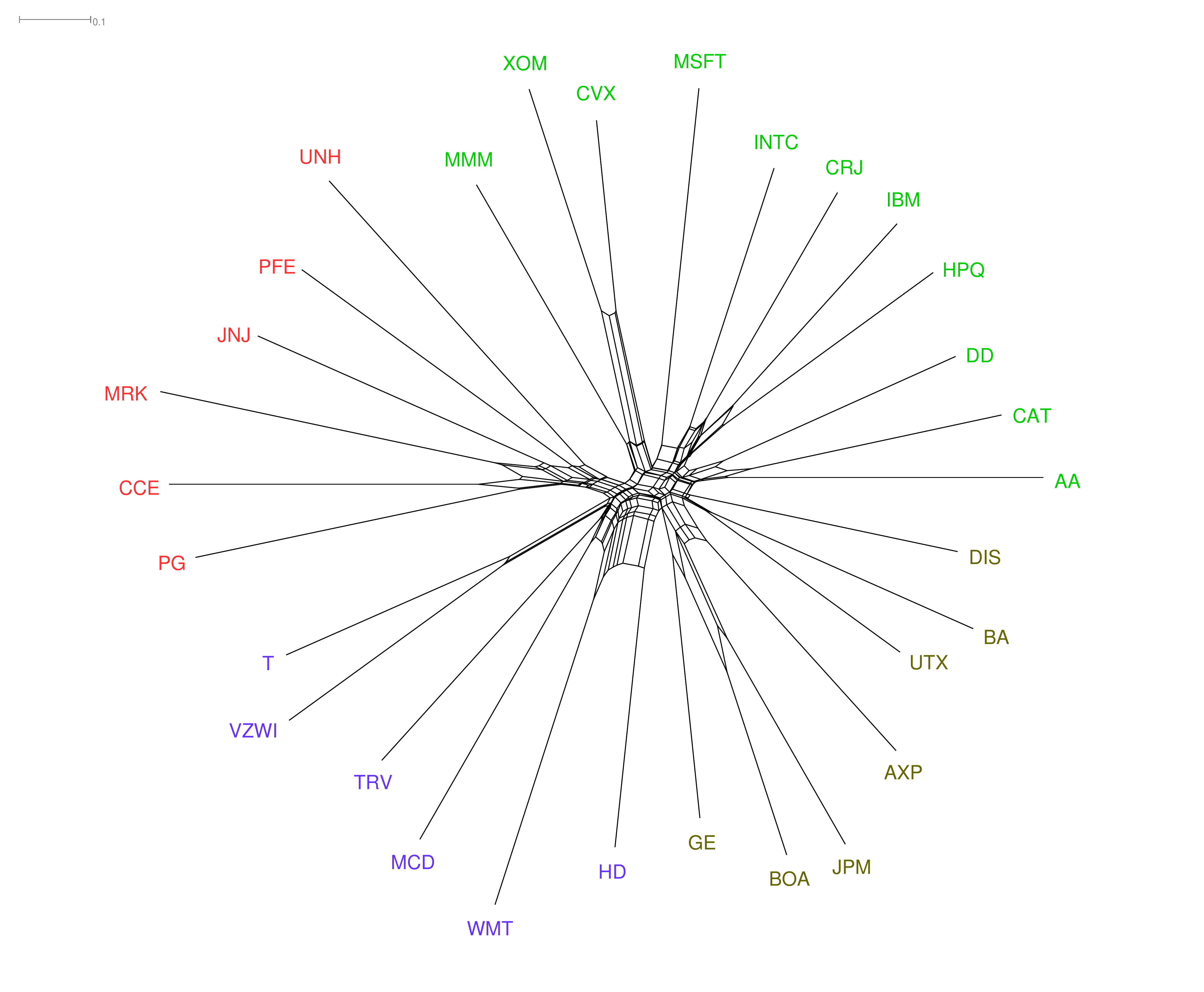}
  \caption{Neighbor-Net splits graph for  Period 3 DJIA with four clusters. 
The cluster
sizes are somewhat unbalanced with clusters of six, six, seven and 11 stocks 
respectively.}
  \label{fig:DJIANNP5Cl4}
\end{figure}

As indicated in the main body of this paper, the circular ordering of
the stocks allows some flexibility in choosing the clusters. In Figures
(\ref{fig:DJIANNP5Cl2}) and (\ref{fig:DJIANNP5Cl4}) shows as
many as six breaks in the network structure giving the analyst some 
freedom in how they are combined to give the required number of clusters,
in our case two and four. Figure (\ref{fig:DJIANNP5Cl2}) shows one possible
way which yields a pair of clusters very similar in size. If one
wished to combine two pairs of the four clusters in Figure  
(\ref{fig:DJIANNP5Cl4}) to form two clusters the most obvious choice is
to group the stocks from T to DIS (reading counter-clockwise) into one
cluster and AA to PG into a second cluster. This would yield clusters of
sizes 13 and 17 stocks respectively.

\clearpage

\section{Extra Results}\label{sec:extraresults}

Table (\ref{tab:period1}) presents the results of the simulations for
groups determined by Period 1 data and tested against in-sample
(Period 1) and out-of-sample (Period 2) data.

Table (\ref{tab:period6}) presents the results of the simulations for
groups determined by Period 3 data and tested against in-sample
(Period 3) and out-of-sample (Period 4) data.

The other results are in the main body of the paper. However,
some of the results
here are discussed in the main body of the paper. 

\begin{table}
\begin{tabular}{lllllll}
Period 1           &        &              &     &         &  & \\
                   &        &              &     &         & Industry & Levene\\
Simulation results & Random & N-Net        & HCT & MST     &  Group & p-value \\
\hline
Mean return \\
(2-stock portfolios) & 26.39  & 23.98     & 29.59 &  31.22 & 24.20 \\
(4-stock portfolios) & 26.95  & 22.14     & 30.00 &  29.59 & 24.31 \\
(8-stock portfolios) & 26.19  & 23.16     & 30.20 &  28.88 & 24.60 \\
\hline
Standard Deviation \\
(2-stock portfolios) & 32.44 & 30.12      & 28.91 & 32.57 & 31.22 & 0.005\\
(4-stock portfolios) & 22.47 & 20.93      & 19.25 & 19.83 & 21.89 & 0.020\\
(8-stock portfolios) & 14.40 & 14.30      & 13.11 & 13.01 & 13.91 & 0.002\\
\hline
Sharpe Ratios \\
(2-stock portfolios) & 0.72  & 0.70       & 0.91*  & 0.87 & 0.68  \\
(4-stock portfolios) & 1.07  & 0.91       & 1.40*  & 1.34 & 0.97  \\
(8-stock portfolios) & 1.61  & 1.41       & 2.07*  & 1.99 & 1.55  \\
\hline 
\\
Period 2           &        &              &     &         & \\
                   &        &              &     &         & Industry& Levene \\
Simulation results & Random & N-Net        & HCT & MST     &  Group  &p-value\\
\hline
Mean return \\
(2-stock portfolios) & 60.80  & 61.25     & 67.93 &  64.38 & 66.35  \\
(4-stock portfolios) & 61.06  & 59.23     & 67.59 &  63.05 & 68.79 \\
(8-stock portfolios) & 61.65  & 59.27     & 67.13 &  62.60 & 67.90 \\
\hline
Standard Deviation \\
(2-stock portfolios) & 30.57 & 32.20      & 27.42 & 33.23 & 33.32 & $9.9\times 10^{-10}$ \\
(4-stock portfolios) & 21.16 & 21.87      & 20.46 & 23.17 & 21.06 & $0.008$ \\
(8-stock portfolios) & 13.77 & 13.54      & 12.96 & 14.63 & 13.49 & 0.017\\
\hline
Sharpe Ratios \\
(2-stock portfolios) & 1.92 & 1.83       & 2.40*  & 1.87 & 1.92  \\
(4-stock portfolios) & 2.78 & 2.60       & 3.20*  & 2.63 & 3.16  \\
(8-stock portfolios) & 4.32 & 4.21       & 5.01*  & 4.13 & 4.87  \\
\hline
\end{tabular}
\caption{Mean portfolio returns and the standard deviations of 1000 replications
of five different portfolio selection methods for in-sample (period 1) and
out-of-sample (period 2). The Sharpe ratios used
a period return of 3.0\% and 2.2\% for periods 1 and 2 respectively. 
The highest Sharpe ratios within each
portfolio size are marked with an
asterisk. The Levene test p-values exclude the HCT because all results
which included the HCT were very highly signficant.}\label{tab:period1}
\end{table}

\begin{table}
\begin{tabular}{lllllll}
Period 3           &        &              &     &         &         &\\
                   &        &              &     &         & Industry& Levene \\
Simulation results & Random & Neighbor-Net & HCT & MST     &  Group & p-value \\
\hline
Mean return \\
(2-stock portfolios) & 24.72  & 27.31     & 36.32 &  24.43 & 24.30  \\
(4-stock portfolios) & 24.87  & 25.69     & 35.54 &  23.77 & 25.63 \\
(8-stock portfolios) & 25.28  & 26.66     & 35.30 &  23.77 & 25.10 \\
\hline
Standard Deviation \\
(2-stock portfolios) & 23.52 & 22.30      & 21.86 & 21.99 & 22.58 & 0.109 \\
(4-stock portfolios) & 16.13 & 15.42      & 14.47 & 15.84 & 14.68 & 0.019\\
(8-stock portfolios) &  9.98 &  8.98      &  8.44 & 10.26 &  9.40 & $3.16\times 10^{-5}$\\
\hline
Sharpe Ratios \\
(2-stock portfolios) & 0.86 & 1.03       & 1.46*  & 0.91 & 0.88  \\
(4-stock portfolios) & 1.27 & 1.36       & 2.15*  & 1.22 & 1.44  \\
(8-stock portfolios) & 2.09 & 2.48       & 3.68*  & 1.89 & 2.20  \\
\hline 
Period 4           &        &              &     &         &  & \\
                   &        &              &     &         & Industry& Levene \\
Simulation results & Random & Neighbor-Net & HCT & MST     &  Group  &p-value\\
\hline
Mean return \\
(2-stock portfolios) & 104.51  & 115.60    & 121.98 &  93.84 & 106.37 \\
(4-stock portfolios) & 104.35  & 115.29    & 121.89 &  96.31 & 107.04\\
(8-stock portfolios) & 105.86  & 116.03    & 120.67 &  96.24 & 107.63\\
\hline
Standard Deviation \\
(2-stock portfolios) & 49.38 & 44.33      & 50.43 & 52.17 & 50.40 & $5.27\times 10^{-5}$\\
(4-stock portfolios) & 34.15 & 30.32      & 33.57 & 35.35 & 34.68 &0.0002\\
(8-stock portfolios) & 22.34 & 19.74      & 19.62 & 22.18 & 22.60 &$ 6.55\times 10^{-5}$\\
\hline
Sharpe Ratios \\
(2-stock portfolios) & 2.08 & 2.57       & 2.39*  & 1.61 & 2.07  \\
(4-stock portfolios) & 3.01 & 3.74*      & 3.58   & 2.68 & 3.04  \\
(8-stock portfolios) & 4.66 & 5.79       & 6.06*  & 4.26 & 4.69  \\
\hline
\end{tabular}
\caption{Mean portfolio returns and the standard deviations of 1000 replications
of five different portfolio selection methods for in-sample (period 3) and
out-of-sample (period 4). The Sharpe ratios used
a period return of  4.4\% and 1.7\% for periods 3 and 4 respectively. 
The highest Sharpe ratios within each
portfolio size are marked with an
asterisk.}\label{tab:period6}
\end{table}

\clearpage

\section{Stocks, Ticker Symbols and Industry Groups}\label{app:stockcodes}

Table (\ref{tab:DJStocks}) represents the 30 stocks in our sample in 
alphabetical order, with their exchanged-assigned ticket symbols and
the industry group.

Table (\ref{tab:DJStocks4Group}) has the stocks divided into the 
four super-groups which we used in the simulations.

\begin{table}[h]
\begin{tabular}{lll}
Company Name  & Symbol &  Industry Group \\
\hline
3M                      &  MMM  &  Diversified Industrials\\
Alcoa                   &  AA   &  Aluminium \\
American Express        &  AXP  &  Insurance and Finance \\
AT\&T                   &  T    &  Telecom \\
Bank of America         &  BOA  &  Insurance and Finance \\
Boeing                  &  BA   &  Aerospace \\
Caterpillar             &  CAT  &  Commercial Vehicles \& Trucks\\
Chevron                 &  CVX  &  Integrated Oil \& Gas \\
Cisco Systems           &  CRJ  &  Telecom \\
Coca Cola               &  CCE  &  Soft Drinks \\
E I Du Pont de Nemours  &  DD   &  Commodity Chemicals \\
Exxon Mobil             &  XOM  &  Integrated Oil \& Gas \\
General Electric        &  GE   &  Diversified Industrial \\
Hewlett-Packard         &  HPQ  &  Computer Hardware \\
Home Depot              &  HD   &  Home Improvement Retailers \\
Intel                   &  INTC &  Computer Hardware \\
International Bus.Mchs. &  IBM  &  Computer Services\\
Johnson \& Johnson      &  JNJ  &  Healthcare \\
JP Morgan Chase \& Co.  &  JPM  &  Insurance and Finance \\
McDonalds               &  MCD  &  Resaurants \& Bars \\
Merck \& Co.            &  MRK  &  Pharmaceuticals \\
Microsoft               &  MSFT &  Technology \\
Pfizer                  &  PFE  &  Pharmaceuticals \\
Procter \& Gamble       &  PG   &  Nondurable Household Products \\
Travelers Cos.          &  TRV  &  Insurance and Finance \\
United Technologies     &  UTX  &  Aerospace \\
United Health GP.       &  UNH  &  Healthcare \\
Verizon Communications  &  VZWI &  Telecom \\
Wal Mart Stores         &  WMT  &  Retailers \\
Walt Disney             &  DIS  &  Broadcasting and Entertainment\\
\hline
\end{tabular}
\caption{The 30 stocks in the Dow Jones sample with their ticker symbols
and industry group.}\label{tab:DJStocks}
\end{table}

\begin{table}
\begin{tabular}{lll}
Company Name  & Symbol &  Industry Group \\
\hline
American Express        &  AXP  &  Insurance and Finance \\
Bank of America         &  BOA  &  Insurance and Finance \\
Johnson \& Johnson      &  JNJ  &  Healthcare \\
JP Morgan Chase \& Co.  &  JPM  &  Insurance and Finance \\
Merck \& Co.            &  MRK  &  Pharmaceuticals \\
Pfizer                  &  PFE  &  Pharmaceuticals \\
Travelers Cos.          &  TRV  &  Insurance and Finance \\
United Health GP.       &  UNH  &  Healthcare \\
\hline
Alcoa                   &  AA   &  Aluminium \\
Boeing                  &  BA   &  Aerospace \\
Chevron                 &  CVX  &  Integrated Oil \& Gas \\
Exxon Mobil             &  XOM  &  Integrated Oil \& Gas \\
United Technologies     &  UTX  &  Aerospace \\
\hline
3M                      &  MMM  &  Diversified Industrials\\
Caterpillar             &  CAT  &  Commercial Vehicles \& Trucks\\
Coca Cola               &  CCE  &  Soft Drinks \\
E I Du Pont de Nemours  &  DD   &  Commodity Chemicals \\
General Electric        &  GE   &  Diversified Industrial \\
Home Depot              &  HD   &  Home Improvement Retailers \\
McDonalds               &  MCD  &  Resaurants \& Bars \\
Wal Mart Stores         &  WMT  &  Retailers \\
Walt Disney             &  DIS  &  Broadcasting and Entertainment\\
\hline
AT\&T                   &  T    &  Telecom \\
Cisco Systems           &  CRJ  &  Telecom \\
Hewlett-Packard         &  HPQ  &  Computer Hardware \\
Intel                   &  INTC &  Computer Hardware \\
International Bus.Mchs. &  IBM  &  Computer Services\\
Microsoft               &  MSFT &  Technology \\
Procter \& Gamble       &  PG   &  Nondurable Household Products \\
Verizon Communications  &  VZWI &  Telecom \\
\hline
\end{tabular}
\caption{The 30 stocks in the Dow Jones sample with their ticker symbols
and industry group with horizontal lines dividing the stocks into
their four super-groups for the portfolio simulations using industry
selection.}\label{tab:DJStocks4Group}
\end{table}

\end{document}